\newif\ifShowKeys
\ShowKeystrue
\ShowKeysfalse

\documentclass[11pt,a4paper]{article} 
\usepackage[height=8.8in,width=6.45in]{geometry}

\usepackage[hidelinks]{hyperref}

\usepackage{tocloft}
\usepackage{sidecap}

\ifShowKeys \usepackage[notcite]{showkeys} \fi

\usepackage{amsmath, amssymb,amsthm}
\usepackage{amsthm}
\numberwithin{equation}{section}

\usepackage{bm,environ,mathrsfs,array,arydshln}
\usepackage{booktabs,float,slashed,hyperref}
\usepackage[mathcal]{euscript}
\usepackage{tensor} 						% Ratcliffe package to write tensors
\usepackage{mathabx}
\usepackage{simpler-wick}

\usepackage{aurical}
\usepackage[T1]{fontenc}

\usepackage[nodayofweek]{date time}
\usepackage{graphicx,epsfig,epic}
\usepackage{tikz}
\usetikzlibrary{decorations.pathreplacing,decorations.markings,snakes,arrows.meta,shapes.misc}
\tikzset{middlearrow/.style={decoration={markings, mark= at position 0.5 with {\arrow{#1}} ,
}, postaction={decorate}}}
\tikzset{cross/.style={cross out , draw=black,  fill opacity = 1 , draw opacity=1 , minimum size=2*(#1-\pgflinewidth), inner sep=0pt, outer sep=0pt},
  cross/.default={1pt}}
\usepackage{subcaption}

\usepackage{framed}						% for shaded equations \begin{shaded}...\end{shaded}
\definecolor{shadecolor}{rgb}{0.95,0.95,0.97}

\usepackage{comment}

\allowdisplaybreaks

\newcommand{\bs}{\begin{shaded}}
\newcommand{\es}{\end{shaded}}
\def\ba#1\ea{\begin{align}#1\end{align}}		% very clever way to bypass the known problem...
\newcommand{\be}{\begin{equation}}
\newcommand{\ee}{\end{equation}}
\newcommand{\mc}{\mathcal }

\newcommand{\la}{\label}

\newcommand{\lp}{\notag \\ & }

\DeclareMathOperator{\tr}{\text{tr}}

\newcommand{\cf}{\textit{cf.} }
\newcommand{\ie}{\textit{i.e.} }

\makeatletter
\DeclareFontFamily{OMX}{MnSymbolE}{}
\DeclareSymbolFont{MnLargeSymbols}{OMX}{MnSymbolE}{m}{n}
\SetSymbolFont{MnLargeSymbols}{bold}{OMX}{MnSymbolE}{b}{n}
\DeclareFontShape{OMX}{MnSymbolE}{m}{n}{
<-6>  MnSymbolE5
   <6-7>  MnSymbolE6
   <7-8>  MnSymbolE7
   <8-9>  MnSymbolE8
   <9-10> MnSymbolE9
  <10-12> MnSymbolE10
  <12->   MnSymbolE12
}{}
\DeclareFontShape{OMX}{MnSymbolE}{b}{n}{
<-6>  MnSymbolE-Bold5
   <6-7>  MnSymbolE-Bold6
   <7-8>  MnSymbolE-Bold7
   <8-9>  MnSymbolE-Bold8
   <9-10> MnSymbolE-Bold9
  <10-12> MnSymbolE-Bold10
  <12->   MnSymbolE-Bold12
}{}

\let\llangle\@undefined
\let\rrangle\@undefined
\DeclareMathDelimiter{\llangle}{\mathopen}%
 {MnLargeSymbols}{'164}{MnLargeSymbols}{'164}
\DeclareMathDelimiter{\rrangle}{\mathclose}%
 {MnLargeSymbols}{'171}{MnLargeSymbols}{'171}
\makeatother

\catcode`,\active

\catcode`\,12

\def\XXint#1#2#3{{\setbox0=\hbox{$#1{#2#3}{\int}$}
     \vcenter{\hbox{$#2#3$}}\kern-.5\wd0}}

\newcommand{\gs}{g_{\text{s}}}

\newcommand{\vev}[1]{\left\langle  #1 \right\rangle}

\newcommand{\abs}[1]{|#1|}
\newcommand{\ratio}[1]{\frac{\vev{\mc W^{#1}}_{c}}{\vev{\mc W}^{#1}}}

\newcommand{\LT}{\stackrel{T\gg 1}{=}}

\DeclareMathOperator{\Res}{Res}

\definecolor{darkred}{rgb}{.8,.1,.01}
\definecolor{darkcyan}{RGB}{57,131,202}
\definecolor{darkblue}{RGB}{59,16,115}

\begin{document}

\begin{titlepage}

%\date{\currenttime}
%\begin{flushright}\boxed{\small{\tt \today \ \ - \ \  \currenttime }}\end{flushright}
%
%\begin{tabbing}
%\hspace*{11.5cm} \=  \kill % set the tabbings
%\> additional information:  \\
%\> none
%\end{tabbing}

\vspace*{15mm}
\begin{center}
{\LARGE\sc  On topological recursion for Wilson loops}\vskip 5pt
{\LARGE\sc  in  $\mc N=4$ SYM at strong coupling}

\vspace*{10mm}

{\Large  M. Beccaria${}^{\,a}$ and A. Hasan${}^{\,a}$}

\vspace*{4mm}
	
${}^a$ Universit\`a del Salento, Dipartimento di Matematica e Fisica \textit{Ennio De Giorgi},\\ 
		and I.N.F.N. - sezione di Lecce, Via Arnesano, I-73100 Lecce, Italy
			\vskip 0.3cm
			
\vskip 0.2cm
	{\small
		E-mail:
		\texttt{matteo.beccaria@le.infn.it, ahasan@gradcenter.cuny.edu}
	}
\vspace*{0.8cm}
\end{center}

\begin{abstract} 
We consider $U(N)$ $\mc N=4$ super Yang-Mills theory and discuss how to 
extract the strong coupling limit of non-planar corrections to  observables involving the $\frac{1}{2}$-BPS 
Wilson loop. Our approach is based on a suitable saddle point treatment of the Eynard-Orantin topological 
recursion in the Gaussian matrix model.
Working directly at strong coupling we 
avoid the usual procedure of  first computing  observables at finite planar coupling $\lambda$,  
order by order in $1/N$, and then taking the $\lambda\gg 1$ limit.
In the proposed approach,  matrix model multi-point resolvents take a simplified form and 
some structures of the genus expansion, 
hardly visible at low order, may be identified and rigorously proved. 
As a sample application, we consider the 
expectation value of multiple coincident circular supersymmetric Wilson loops as well as  their
correlator with single trace chiral operators. For these quantities 
we provide novel results about the structure of their  genus expansion at large tension, generalising 
 recent results in \href{https://arxiv.org/abs/2011.02885}{\color{darkblue}{arXiv:2011.02885}}.
\end{abstract}
\end{titlepage}

\tableofcontents
\vspace{1cm}

\section{Introduction and results}

The recent papers \cite{Giombi:2020mhz,Beccaria:2020ykg,Beccaria:2021ksw} focused on certain features of
higher genus corrections to BPS Wilson loops in dual theories related by AdS/CFT. By means of  supersymmetric localization,
gauge theory predictions are  available as matrix model integrals that depend non-trivially
on the number of colours $N$ and 't Hooft planar coupling $\lambda$ (mass deformations will not be relevant here). 
The large $N$ expansion may be computed at high order  starting from exact expressions in the matrix model
or by perturbative loop equation methods, like topological recursion \cite{Eynard:2007kz}. On the string side, the gauge theory 
parameters $N, \lambda$ may be replaced by the string coupling $\gs$ and  tension $T$. World-sheet genus expansion is 
a natural perturbation theory controlled by powers of $\gs$ accompanied by corrections in inverse string tension, \ie
$\sigma$-model quantum corrections. The two expansions are  expected to match according to AdS/CFT, but practical
tests are of course non-trivial. On the gauge side, a rich set of predictions is obtained extracting the dominant strong coupling corrections order by order in $1/N$, \ie well beyond planar level.
On string side, this should reproduce the large tension limit $T\gg 1$ at specific genera, whose
independent determination is obviously very hard beyond leading order. In spite of that, 
one can still look at manifestations of its expected structural properties in the $1/N$ gauge theory expansion.

The simplest example where this strategy may be concretely illustrated is the expectation value $\vev{\mc W}$
of the $\frac{1}{2}$-BPS circular Wilson loop in $U(N)$ $\mc N=4$ SYM. The expression for  $\vev{\mc W}$  
is known at finite $N$ and $\lambda = N\,g^{2}_{\rm YM}$ exactly \cite{Erickson:2000af,Drukker:2000rr,Pestun:2007rz,Zarembo:2016bbk} and is given
by  the Hermitian  Gaussian  one-matrix model average
\be
\la{1.1}
\vev{\mc W} = %\vev{\tr e^{M}} = 
\int \mc DM\, \tr e^{\frac{\sqrt\lambda}{2}\,M} e^{-\frac{N}{2}\tr M^{2}} = e^{\frac{\lambda}{8N}}\,L_{N-1}^{1}\bigg(-\frac{\lambda}{4N}\bigg).
\ee
%where 
%the matrix measure in terms of  the $M$ eigenvalues $\lambda_{1}, \dots \lambda_{N}$ reads 
%$\mc DM = \Delta(\lambda)^{2}d^{N}\lambda$ and the Vandermonde factor is $\Delta(\lambda) = \prod_{1\le i<j\le N}(\lambda_{i}-\lambda_{j})$.
%the normalization $\mc Z$ is such that $\vev{1}=1$.
In this case, the relation among the gauge theory parameters $\lambda, N$ and $\gs, T$ in the  dual $\text{AdS}_{5}\times S^{5}$ IIB superstring is  \cite{Maldacena:1997re}
\be
\gs = \frac{\lambda}{4\pi N}, \qquad T=\frac{\sqrt\lambda}{2\pi}.
\ee
At large tension, (\ref{1.1}) takes the following form
\ba
\la{1.3}
\langle \mc W\rangle &=    \frac{1}{2\pi}\,\frac{\sqrt T}{\gs}\,e^{2\pi T+\frac{\pi}{12}\frac{\gs^{2}}{T}}\,\bigg[1+\mc O(T^{-1})\bigg]
\LT e^{2\pi\,T}\, \mathsf{f}\bigg(\frac{\gs^{2}}{T}\bigg), \\
\la{1.4}
\mathsf{f}(x) &= x^{-1/2}\,\exp\bigg(\frac{\pi}{12}\,x^{2}\bigg).
\ea
The structure of (\ref{1.3}) is consistent with the dual representation of the Wilson loop expectation value 
as the string path integral over world-sheets ending on a circle at $\partial\text{AdS}$. \footnote{
The  exponential factor  $\exp(2\pi T)$ comes from 
the  AdS$_{2}$ minimal surface \cite{Berenstein:1998ij,Drukker:1999zq,Drukker:2000ep}. Upon expansion  in $\gs$, the power of the string coupling 
is minus the Euler number  of a disc with $p$  handles ($\chi=1-2p$). The fact that each power of $\gs$ is accompanied at large tension by 
a factor $1/\sqrt{T}$ is non-trivial and explained in \cite{Giombi:2020mhz}. A similar structure holds for Wilson loops in ABJM theory, dual to 
string on $\text{AdS}_{4}\times \mathbb{CP}^{3}$.}

A similar large tension analysis is presented in \cite{Beccaria:2020ykg} for other quantities related again to the $\frac{1}{2}$-BPS Wilson loop in $\mc N=4$ SYM. 
In particular, 
one can consider the normalised ratio of $n$  coincident Wilson loops. 
\footnote{See \cite{Johnson:2021owr} for a  recent application of such coincident loops in matrix models associated with JT gravity.}
This requires  consideration of matrix integrals which are generalisations of (\ref{1.1}), but  whose $1/N$ expansion is much 
more difficult to extract. \footnote{Indeed, in this case one does not have a simple result like (\ref{1.1}), but instead multiple finite sum of $\sim N$ terms, see 
for instance Eq.~(4.3) in \cite{Beccaria:2020ykg} for $n=2$.}
The semiclassical exponential factors $\sim e^{2\pi T}$ cancel
and the ratio $\vev{\mc W^{n}}/\vev{\mc W}^{n}$ is again organised in powers of $\gs^{2}/T$, \cf (\ref{1.3}),
\ba
\la{1.5}
\frac{\vev{\mc W^{n}}}{\vev{\mc W}^{n}} &\LT \mathsf{W}_{n}\bigg(\frac{\pi\gs^{2}}{T}\bigg), 
\ea
where the first three terms of the scaling function $\mathsf{W}_{n}$ have been computed in  \cite{Beccaria:2020ykg} and read
\ba
\mathsf{W}_{n}(x) &=  1+\frac{n\,(n-1)}{2}\,x+\frac{n\,(n-1)\,(3n-5)\,(n+2)}{24}\,x^{2}\lp
+\frac{n\,(n-1)\,(15n^{4}+30n^{3}-75n^{2}-610 n+1064)}{720}\,x^{3}+\cdots.
\ea
A third example of scaling functions emerging in the large tension limit are 
normalised correlators of $\mc W$ with a single trace chiral operator $\mc O_{J}\sim \tr \Phi^{J}$ 
\cite{Berenstein:1998ij,Semenoff:2001xp}
recently reconsidered in \cite{Beccaria:2020ykg}. In this case,
the large tension limit is characterised by a  different scaling combination 
\be
\la{1.7}
\frac{\vev{\mc W\,\mc O_{J}}}{\vev{\mc W}} \LT  J\,\bigg(\frac{\pi}{2}\bigg)^{J/2}\,T\,\mathsf{F}_{J}\bigg(\frac{\gs^{2}}{T^{2}}\bigg), \qquad 
\mathsf{F}_{J}(x) = \frac{2}{J\,\sqrt{x}}\sinh\bigg(J\,\text{arcsinh}\frac{\sqrt x}{2}\bigg),
\ee
where we draw attention to the non-trivial dependence of $\mathsf{F}_{J}(x)$ on the R-charge $J$.  \footnote{Through an analytical continuation 
it is possible to capture $\mathsf{F}_{J}$ by a D3-brane calculation, see \cite{Giombi:2006de}.}
%
%it is interesting to recall that the function $\mathsf{F}_{J}$  admits a ``D3-brane'' interpretation \cite{Giombi:2006de}.  \footnote{This is non trivial already in the case of 
%the simple Wilson loop. Indeed one has to consider the  $k$-symmetric representation and take the limit $k,N\to\infty$ with fixed $\kappa=\frac{k\gs}{2T}$. This gives 
%$\vev{\mc W} = \exp(-S_{\rm D3})$ where $S_{\rm D3}= N\,f(\kappa)$ ia the D3-brane action \cite{Drukker:2005kx}. Taking back $k\to 1$ matches the large tension expansion 
%of $\vev{\mc W}$. A similar treatment may be repeated for (\ref{1.7}), see  Eq.~(4.20) in \cite{Giombi:2006de}.}
%

Beyond  proving  general structures as in (\ref{1.3}), (\ref{1.5}), and (\ref{1.7}),  
it is important to develop methods to determine the detailed form of  scaling functions
like  $\mathsf{f}$, $\mathsf{W}_{n}$ and $\mathsf{F}_{J}$. A common approach is to 
compute the $1/N$ expansion at finite planar coupling $\lambda$ in the Hermitian Gaussian one-matrix model, and then take the strong coupling limit $\lambda\gg 1$. 
For instance, in the case of  $\vev{\mc W}$, one has the exact representation  at finite $\lambda$ \cite{Okuyama:2006ir}
\be
\la{1.8}
\vev{\mc W}  = \frac{2N}{\sqrt\lambda}\,\mathop{\text{Res}}_{x=0} \Big[
\, e^{\frac{\lambda}{4N}\,H(\frac{\sqrt\lambda}{4N}\,x)}\sum_{n=0}^{\infty}I_{n}(\sqrt\lambda)\,x^{-n}\Big]\ ,
\qquad \quad  H(x) \equiv  \frac{1}{2}\Big(\coth x-\frac{1}{x}\Big).
\ee
From (\ref{1.8}), we get all coefficients of the $1/N$ power series in terms of  explicit combinations of modified Bessel functions ($I_{n}\equiv I_{n}(\sqrt\lambda)$), see also \cite{Drukker:2000rr}, 
\ba
\la{1.9}
\vev{\mc W} =  \frac{2N\, I_1}{\sqrt{\lambda }}& +\frac{\lambda  I_2}{48 N}+\frac{1}{N^{3}}\Big(\frac{\lambda ^{5/2} I_3}{9216}-\frac{\lambda ^2 
I_4}{11520}\Big)+\frac{1}{N^{5}}\Big(\frac{\lambda ^4 I_4}{2654208}-\frac{\lambda 
^{7/2} I_5}{1105920}+\frac{\lambda ^3 
I_6}{1935360}\Big)+\cdots.
%\lp
%+\frac{1}{N^{8}}\Big(\frac{\lambda ^{11/2} 
%I_5}{1019215872}-\frac{\lambda ^5 I_6}{212336640}+\frac{\lambda 
%^{9/2} I_7}{137625600}-\frac{\lambda ^4 
%I_8}{309657600}\Big)+\cdots.
%\lp
%+\frac{1}{N^{10}}\Big(\frac{\lambda ^7 
%I_6}{489223618560}-\frac{\lambda ^{13/2} I_7}{61152952320}+\frac{17 
%\lambda ^6 I_8}{356725555200}-\frac{\lambda ^{11/2} 
%I_9}{17836277760}+\frac{\lambda ^5 
%I_{10}}{49049763840}\Big)+\cdots
\ea
When each term of this expression is expanded at large $\lambda$, the result takes the 
simple exponential form (\ref{1.3}). Of course,  the case of $\vev{\mc W}$ is particularly simple because of the compact closed formula (\ref{1.1})
leading to (\ref{1.8}).  Somehow, a similar situation occurs in the case of the scaling function $\mathsf{F}_{J}$ in (\ref{1.7}). 
Indeed, the correlator $\vev{\mc W\,\mc O_{J}}$ admits the  representation \cite{Okuyama:2006jc}
\be
\la{1.10}
\vev{\mc W\, \mc O_{J}} = \bigg(\frac{2}{\pi}\bigg)^{1-J/2}\,\frac{N^{2}}{\sqrt\lambda}\,e^{\frac{\lambda}{8N}}\oint\frac{dz}{2\pi i}
z^{J}\,e^{\frac{\sqrt\lambda}{2}z}\,\Big(1+\frac{\sqrt\lambda}{2N z}\Big)^{N}\,
\Big[\Big(1+\frac{\sqrt\lambda}{2N z}\Big)^{J}-1\Big],
\ee
and  one can prove (\ref{1.7}) from this formula, which is exact at finite $N$ and $\lambda$ \cite{Giombi:2006de,Beccaria:2020ykg}. 

However, as soon as the observables under study become more complicated, it is increasingly difficult to extract  the genus expansion  order by order in $1/N$ at finite $\lambda$.
An example are multiple coincident Wilson loops $\vev{\mc W^{n}}$ -- not to be confused with multiply wound loops -- or multi-trace chiral operators \cite{Aprile:2020uxk}. In this case,
exact expressions are not available or are too cumbersome to be useful. 
Toda recursion relations \cite{Gerasimov:1990is,Morozov:1994hh,Morozov:1995pb,Mironov:2005qn,Morozov:2009uy} are 
a possible method to determine the $1/N$ expansion, but work well only for simple observables \cite{Beccaria:2020ykg} (and their scope is limited to the Gaussian matrix model).
A more general approach is to take advantage of topological recursion \cite{Eynard:2004mh,Eynard:2008we} which is an efficient way to organise the 
hierarchy  the matrix model  loop equations. \footnote{See also 
\cite{Chester:2019jas,Chester:2019pvm,Binder:2019jwn,Chester:2020dja,Chester:2020vyz}
for other recent applications of topological recursion to $\mc N=4$ SYM.}
In practice, a serious bottleneck in applying this method is the rapid increase of 
 computational complexity at higher  genus, see for instance \cite{Okuyama:2018aij}. For these reasons, it seems important to devise a 
version of topological recursion suitable for strong coupling directly.

In this paper, we take a first step in this direction. We illustrate a practical approach to work out  topological recursion at strong coupling by isolating 
dominant contributions at large tension.
Despite its simplicity, the method turns out to be rather effective. 
As an illustration, we present an algorithm for computing the function $\mathsf{W}_{n}(x)$ in (\ref{1.5}) at any desired order with minor effort, and 
we illustrate remarkable exponentiation properties of the dominant terms at large $n$. This result will be cross
checked by means of an extension to all $n$ of the Toda recursion method used in \cite{Beccaria:2020ykg} for $n=2,3$. As a second application, we shall 
prove that the structure of (\ref{1.7}) is rather special and does not extend to the normalized correlators of a chiral primary single trace operator with multiple 
coinciding Wilson loops, \ie ratios
$\vev{\mc W^{n}\mc O_{J}}/\vev{\mc W^{n}}$ when $n>1$. Instead, we prove that the relevant scaling variable is $\gs^{2}/T$ and that the dependence on the R-charge is
\be
\la{1.11}
\frac{\vev{\mc W^{n}\,\mc O_{J}}}{\vev{\mc W^{n}}} \LT J\,\bigg(\frac{\pi}{2}\bigg)^{J/2}\,n\,\bigg[T+(J^{2}-1)\,\mathsf{H}_{n}\bigg(\frac{\pi\gs^{2}}{T}\bigg)\bigg],
\ee
where the function $\mathsf H_{n}(x)$ is independent of $J$ and may be computed in terms of $\mathsf{W}_{n}$ by the relation 
\be
\la{1.12}
\mathsf{H}_{n}(x) =  \frac{x}{2\pi}\,\bigg[\frac{1}{12}+\frac{1}{n}\,(\log \mathsf{W}_{n}(x))'\bigg].
\ee
The derivation of these results is  straightforward in the framework of the strong coupling version of topological recursion, and far from trivial by other methods. 
A similar approach is expected to be useful and apply in harder cases with separated Wilson loops or more local operator insertions. 
Some of these problems can be mapped to multi-matrix models calculations \cite{Giombi:2012ep} that would be interesting to 
study by a suitable strong coupling limit  of more general topological recursions  \cite{Eynard:2007nq}.

\vskip 10pt
The detailed plan of the paper is as follows. In Section \ref{sec:top-rec} we briefly recall the structure of topological recursion for $\mc N=4$ SYM
and its application to the evaluation of $\vev{\mc W}$.
In Section \ref{sec:saddle} we show how to perform a saddle point expansions at strong coupling in the considered problems. We clarify what are the relevant features
of resolvents in that regime. Section \ref{sec:topological_recursion_strong} presents the  strong coupling version of topological recursion, 
capturing the reduced resolvents.
In Section \ref{sec:coincW} we apply this formalism to our first application, \ie the computation of $\vev{\mc W^{n}}$ at large tension. 
In Section \ref{sec:toda}, as a non-trivial check of our approach,  the same results are obtained by solving in the strong coupling limit a suitable Toda recursion for
correlators of traced exponentials in the Gaussian matrix model. 
Finally, in Section \ref{sec:chiral} we discuss the 
correlators $\vev{\mc W^{n}\mc O_{J}}$ between coincident Wilson loops and a single trace chiral operator. The relation 
with the scaling function characterising $\vev{\mc W^{n}}$ is proved in Section \ref{sec:H-W}.

\section{Topological recursion for the Gaussian Matrix Model}
\la{sec:top-rec}

For a Hermitian one-matrix model with potential  $V$, the spectral curve is defined by \cite{Eynard:2008we,Eynard:2015aea}
\be
\la{2.1}
y^{2}-\frac{1}{2}V'(x)\,y+P(x)=0,\qquad P(x) = \frac{1}{N}\vev{\tr\frac{V'(x)-V'(M)}{x-M}},
\ee
where $\langle \mc O(M)\rangle = \int DM\,e^{-N\,\tr V(M)}\,\mc O(M)$ and normalization is fixed by $\vev{1}=1$.
In the Gaussian case, $V(M) = \frac{1}{2}M^{2}$, \cf  (\ref{1.1}),  and the curve (\ref{2.1}) takes the form 
\be
\la{2.2}
y^{2}-xy+1=0,
\ee
admitting the rational (complex) parametrization
\be
\la{2.3}
x=z+\frac{1}{z},\qquad y=\frac{1}{z}.
\ee
The $n$-point resolvent is defined as the connected correlator \footnote{
Connected correlators $\vev{X_{1}X_{2}\cdots}_{c}$ are functional derivatives of the logarithm of the generating function of correlators with respect to sources
coupled to $X_{i}$ operators. 
}
\be
W_{n}(x_{1}, \dots, x_{n}) = \vev{\tr\frac{1}{x_{1}-M}\cdots\tr \frac{1}{x_{n}-M}}_{\rm c},
\ee 
and admits the following genus expansion at large $N$
\be
\la{2.5}
W_{n}(x_{1}, \dots, x_{n}) = \sum_{g=0}^{\infty}\frac{1}{N^{n-2+2g}}\,W_{n,g}(x_{1}, \dots, x_{k}).
\ee
The functions $W_{n}(x_{1}, \dots, x_{n})$ may be traded by multi-differentials on the algebraic curve (\ref{2.2})
\be
\la{2.6}
\omega_{n,g}(z_{1}, \dots, z_{n}) = W_{n,g}(x(z_{1}), \dots, x(z_{n}))\,dx(z_{1})\cdots dx(z_{n})\ .
\ee
Multi-trace connected correlators may be computed as contour integrals around the cut
\be
\la{2.7}
\vev{\prod_{i}\tr \mc O_{i}(M)}_{\rm c} = \sum_{g=0}^{\infty}\frac{1}{N^{n-2+2g}}\ \frac{1}{(2\pi\,i)^{n}}\ \oint \omega_{n,g}(z_{1}, \dots, z_{n})\,\prod_{i}\mc O_{i}(x(z_{i}))\ .
\ee
Higher genus resolvents obey the topological recursion 
% \footnote{Notice that the resolvent in the topological recursion with argument $1/\zeta$ have 
%an extra $-1/\zeta^{2}$ from the implicit differential.}
\ba
\la{2.8}
 \omega_{1,0}(z) &= \frac{1}{z}\left(1-\frac{1}{z^{2}}\right)\,dz, \qquad 
 \omega_{2,0}(z_{1}, z_{2}) = \frac{dz_{1}dz_{2}}{(z_{1}-z_{2})^{2}},  \\
 \omega_{n,g}(z_{1}, \bm{z}) &= \mathop{\text{Res}}_{\zeta=1,-1}\ K(z_{1}, \zeta)\,\bigg[
\omega_{n+1, g-1}(\zeta, \zeta^{-1}, \bm{z})+\sum_{h\le g}\sum_{\bm{w}\subset\bm{z}}\omega_{|\bm{w}|, h+1}(\zeta, \bm{w})\,\omega_{n-|\bm{w}|, g-h}(\zeta^{-1}, \bm{z}\backslash\bm{w})\bigg],\notag \\
 K(z,w) &= \frac{w^{3}}{2\,(w^{2}-1)(z-w)(zw-1)}\frac{dz}{dw}.\notag
\ea
where $\bm{z} = (z_{2}, \dots, z_{n})$,  $\bm{w}$ is a  subset of $\bm{z}$ (preserving the order of the variables),
$|\bm{w}|$ is the number of elements of $\bm{w}$, and $\bm{z}\backslash\bm{w}$ is the complement
of $\bm{w}$ in $\bm{z}$.  In the double sum we exclude the two cases
$(h, \bm{w}) = (0,\emptyset)$ and $(h, \bm{w}) = (g, \bm{z})$.
%Finally, the kernel $K(z,w)$ has the explicit expression 
%\be
%K(z,w) = \frac{w^{3}}{2\,(w^{2}-1)(z-w)(zw-1)}\frac{dz}{dw}.
%\ee
The recursion (\ref{2.8}) allows to compute the following quantities in triangular sequence ( the number under brace is the total weight $g+n$)
\be
\la{2.10}
\underbrace{\omega_{1,1}}_{2}\to \underbrace{\omega_{3,0}\to \omega_{2,1}\to \omega_{1,2}}_{3}\to \underbrace{\omega_{4,0}\to \omega_{3,1}\to \omega_{2,2}\to \omega_{1,3}}_{4}\to \cdots.
\ee
Apart from the seeds $\omega_{1,0}$ and $\omega_{2,0}$, all other resolvents 
have poles in the $z_{i}$ variables only at the special points $\pm 1$.
The first  entries in (\ref{2.10})  read (omitting the $dz_{1}\cdots dz_{n}$ differentials) 
\ba
\la{2.11}
\omega_{1,1}(z) &= \frac{z^{3}}{(z^{2}-1)^{4}}, \notag \\
\omega_{3,0}(z_{1},z_{2},z_{3}) &= -\frac{1}{2\,(z_{1}-1)^{2}(z_{2}-1)^{2}(z_{3}-1)^{2}}+\frac{1}{2\,(z_{1}+1)^{2}(z_{2}+1)^{2}(z_{3}+1)^{2}}, \notag \\
\omega_{2,1}(z_{1},z_{2}) &= \frac{1}{4 (z_{1}^{2}-1)^6 
(z_2^{2}-1)^6}\bigg[
4 z_1^3 z_2 (1+z_2^2){}^2 (1-7 z_2^2+z_2^4)+4 z_1^7 z_2 
(1+z_2^2){}^2 (1-7 z_2^2+z_2^4) \lp
+5 (z_2^4+z_2^6)+5 z_1^{10} 
(z_2^4+z_2^6)+4 z_1 (z_2^3+3 z_2^5+z_2^7)+4 z_1^9 (z_2^3+3 
z_2^5+z_2^7)  \lp
+3 z_1^2 (z_2^2-6 z_2^4-6 z_2^6+z_2^8)+3 z_1^8 (z_2^2-6 
z_2^4-6 z_2^6+z_2^8)  \lp
+12 z_1^5 (z_2-4 z_2^3+16 z_2^5-4 
z_2^7+z_2^9)+z_1^4 (5-18 z_2^2+23 z_2^4+23 z_2^6-18 z_2^8+5 
z_2^{10})\lp 
+z_1^6 (5-18 z_2^2+23 z_2^4+23 z_2^6-18 z_2^8+5 
z_2^{10})
\bigg], \notag \\
\omega_{1,2}(z) &=-\frac{21\,z^{7}\, (1+3 z^2+z^{4})}{(-1+z^2)^{10}},
\ea
and so on. The expression of $\omega_{2,1}$ shows how explicit results become quickly unwieldy. 

\paragraph{Analysis of the simple loop  $\vev{\mc W}$ }

It is useful illustrate how resolvents are used to 
compute the genus expansion of the simple loop expectation value $\vev{\mc W}$. We have 
\ba
\vev{\mc W}  &= \int \mc DM\, \tr e^{\frac{\sqrt\lambda}{2}\,M} e^{-\frac{N}{2}\,\tr M^{2}} \stackrel{N\to \infty}{=}
N\,\sum_{g=0}^{\infty}\frac{1}{N^{2g}}\,\vev{\mc W}_{g},\qquad
\vev{\mc W}_{g} =  \frac{1}{2\pi i}\, \oint \omega_{1,g}(z)\,e^{\frac{\sqrt\lambda}{2}\,(z+1/z)}.
\ea
The leading term is simply \footnote{
We use the generating function $e^{\frac{x}{2}(z+1/z)} = \sum_{n=-\infty}^{\infty}I_{n}(x)\,z^{n}$ and the identity $I_{0}(x)-I_{2}(x) = \frac{2}{x}\,I_{1}(x)$.
}
\be
\la{2.13}
\vev{\mc W}_{0} = \, \oint \frac{dz}{2\pi i}\frac{1}{z}\left(1-\frac{1}{z^{2}}\right)\,e^{\frac{\sqrt\lambda}{2}\,(z+1/z)} = \frac{2}{\sqrt\lambda}\,I_{1}(\sqrt\lambda),
\ee
in agreement with the well known planar result.
The next-to-leading term is 
\be
\vev{\mc W}_{1} =  \oint\frac{dz}{2\pi i} \frac{z^{3}}{(z^{2}-1)^{4}}\,e^{\frac{\sqrt\lambda}{2}\,(z+1/z)}
\ee
The contour encircles all three singular points, but one can check that there are no residues from $z=\pm 1$. Thus, integrating by parts two times gives
\ba
\vev{\mc W}_{1} &= \frac{\lambda}{48}\oint\frac{dz}{2\pi i} \frac{1}{z^{3}}\,e^{\frac{\sqrt\lambda}{2}\,(z+1/z)} = \frac{\lambda}{48}\,I_{2}(\sqrt\lambda),
\ea
which is the well known $1/N^{2}$ correction. A similar manipulation can be repeated for the next order. Integrating by parts five times gives
\ba
\vev{\mc W}_{2} &= \oint\frac{dz}{2\pi i} \frac{21\,z^{7}\, (1+3 z^2+z^{4})}{(z^2-1)^{10}}\,e^{\frac{\sqrt\lambda}{2}\,(z+1/z)} = 
\frac{\lambda^{5/2}}{92160} \oint\frac{dz}{2\pi i} \bigg(\frac{1}{z^{6}}+\frac{9}{z^{4}}\bigg)\,e^{\frac{\sqrt\lambda}{2}\,(z+1/z)} \notag \\
&= \frac{\lambda^{5/2}}{92160}\bigg[I_{5}(\sqrt\lambda)+9\,I_{3}(\sqrt\lambda)\bigg] = 
\bigg[\frac{\lambda ^{5/2} }{9216}\,I_3(\sqrt\lambda)-\frac{\lambda ^2 }{11520}\,I_4(\sqrt\lambda)\bigg],
\ea
in agreement with the $1/N^{3}$ term in (\ref{1.9}).

In the case of $\vev{\mc W}$, this method may be extended to all orders in the $1/N$ expansion, and can also be generalized to give explicit Bessel function combinations
for higher point resolvents at finite $\lambda$, see for instance \cite{Okuyama:2018aij}.  
Nevertheless, the calculation quickly becomes impractical at higher orders due to the very involved expressions that are generated going recursively through the chain of evaluations (\ref{2.10}).
Also, as we explained in the introduction, we are ultimately interested in extracting the large tension limit and want to bypass the cumbersome procedure of first obtaining exact expressions at finite 
$\lambda$, and then expand them at $\lambda\gg 1$. For instance, in the above genus-two contribution both Bessel functions give a similar leading asymptotic contribution  due to  the expansion 
\be
I_{n}(\sqrt\lambda) = \frac{1}{\sqrt{2\pi}}\lambda^{-1/4}\,\bigg(1+\frac{4n^{2}-1}{8}\frac{1}{\sqrt\lambda}+\cdots \bigg)\,e^{\sqrt\lambda}+\cdots, 
\ee
and it would be desirable to pin the total contribution in a more direct way.
To this aim, one needs to study  (\ref{2.8})  working at  strong coupling from the beginning and making more transparent the origin of the dominant terms. 
The next section will be devoted to this problem.

\section{Saddle point methods for Wilson loops}
\la{sec:saddle}

In this section, we discuss how to extract dominant terms from integrals like (\ref{2.13}) by saddle point evaluation. Although this is a fairly well known topic,
we want to emphasize some specific technical issues that are relevant in the calculations we are interested in. To this aim, we consider the large $\sigma\to +\infty$ expansion of a contour integral of the form 
\be
I(\sigma) = \oint dz\, g(z)\, e^{-\sigma\,f(z)}.
\ee
Suppose that $f(z)$ has a critical point $\bar z$ where $f'(\bar z)=0$. Deforming the contour such that it passes through $\bar z$ with constant $\text{Im} f(z)$ along the contour locally around $\bar z$, we write 
($\bar f=f(\bar z)$, $\bar f'' = f''(\bar z)$, $z(0) = \bar z$)
\be
I(\sigma) = e^{-\sigma\,\bar f}\,\int_{-\infty}^{\infty}dt\, \frac{dz}{dt}\, g(z(t))\, e^{-\sigma\frac{1}{2}\,\bar f''\,t^{2}+\cdots}.
\ee
If $g(\bar z)$ is finite, we simply extract it from the integral and perform the Gaussian integral. In the following, we shall be interested in the case 
when $g$ has \underline{an odd zero or an even pole}
around the saddle point. In the case of a zero with 
\be
\la{3.3}
g \stackrel{t\to 0}{=} A\,t^{2m-1}+B\,t^{2m}+\cdots,
\ee
we just include it in the Gaussian integration and get 
\ba
\la{3.4}
I(\sigma) &= e^{-\sigma\,\bar f}\,\int_{-\infty}^{\infty}dt\, [A t^{2m-1}z'(0)+(Bz'(0)+Az''(0))\,t^{2m}+\cdots]\, e^{-\sigma\frac{1}{2}\,\bar f''\,t^{2}+\cdots} \notag \\
&= \sqrt{2\pi}\,[B\,z'(0)+A\,z''(0)]\,e^{-\sigma\,\bar f}\,(2m-1)!!\,(\sigma\,\bar f'')^{-m-\frac{1}{2}}+\cdots.
\ea
In the case of a pole with 
\be
g \stackrel{t\to 0}{=} A\,t^{-2m}+\cdots,
\ee
we compute the finite quantity \footnote{This is equivalent to an implicit integration by parts. In both cases we have to be careful about the poles at $t=0$
  since a non-zero residue for the pole causes a discontinuity in the contour. In our discussion, this will not matter because topological recursion ensures that this residue is always zero, 
  when computing expectation values of  functions of the  matrix model variable. See last section for examples and Appendix \ref{app:toprec-sub} for general details. }
\ba
\frac{d^{m}}{d\sigma^{m}}\bigg[e^{\sigma\,\bar f}\,I(\sigma)\bigg] &= \bigg(-\frac{\bar f''}{2}\bigg)^{m}\,\int_{-\infty}^{\infty}dt\, \frac{dz}{dt}\, g(z(t))\, \,t^{2m}\, e^{-\sigma\frac{1}{2}\,\bar f''\,t^{2}+\cdots}\notag \\
&=  \bigg(-\frac{\bar f''}{2}\bigg)^{m}\, z'(0)\, A\, \sqrt\frac{2\pi}{\sigma\bar f''}+\cdots.
\ea
Integrating back in $\sigma$ gives then 
\be
\la{3.7}
I(\sigma) = \pi\,A\,z'(0)\,e^{-\sigma\,\bar f}\frac{(-1)^{m}}{\Gamma(m+\frac{1}{2})} \bigg(\frac{\sigma\bar f''}{2}\bigg)^{m-\frac{1}{2}}+\cdots. = 
\sqrt{2\pi}\,A\,z'(0)\,e^{-\sigma\,\bar f}\,(-1)^{m}\,\frac{(\sigma\bar f'')^{m-\frac{1}{2}}}{(2m-1)!!}+\cdots.
\ee

\paragraph{Revisiting $\vev{\mc W}$ at strong coupling}

These formulas may be applied to contour integrals involving Wilson loops and higher order resolvents. Let us illustrate this once again in the case of the simple Wilson loop (\ref{1.1}). The planar contribution
in (\ref{2.13}) has $\sigma=\sqrt\lambda$, $f(z) = -\frac{1}{2}(z+1/z)$ and $g(z) = \frac{1}{z}\left(1-\frac{1}{z^{2}}\right)$. The dominant contribution at large $\lambda$ comes
from the saddle point at $z=1$ which is a zero of $g(z)$ of linear order. The parametrization is $z(t) = e^{i t}$ thus $\bar f''=1$. Expanding $g(z)$ around the zero and taking the first even term 
gives (\ref{3.3}) with $A=2i$ and $B=4$ and $m=1$. Evaluation of (\ref{3.4}) gives then 
\be
\vev{\mc W}_{0} = \sqrt\frac{2}{\pi}\lambda^{-3/4}e^{\sqrt\lambda}+\cdots,
\ee
in agreement with (\ref{1.3}). All the higher genus corrections have  even poles at $z=\pm 1$. Again, the leading contribution comes from $z=1$ and may be computed using (\ref{3.7}). For instance, 
at genus one we have
\be
g(z) = \frac{1}{2\pi i}\frac{z^{3}}{(z^{2}-1)^{4}},\qquad g(z(t)) = -\frac{i}{32\pi}\frac{1}{t^{4}}+\cdots \quad \to A = -\frac{i}{32\pi}, \ m=2,
\ee
and 
\be
\vev{\mc W}_{1} = \sqrt{2\pi}\,\frac{-i}{32\pi}\,i\,e^{\sqrt\lambda}\frac{(\sqrt\lambda)^{2-\frac{1}{2}}}{3!!}+\cdots = \frac{\lambda^{3/4}}{48\sqrt{2\pi}}\,e^{\sqrt\lambda}+\cdots.
\ee
Similarly at genus 2 and higher we can check that this procedure reproduces the expansion (\ref{1.3}). Higher order corrections in $1/\sqrt\lambda$ may also be computed in the same way
just by doing Gaussian integration with more accuracy. For instance, we know that (up to exponentially suppressed terms)
\be
\la{3.11}
\vev{\mc W}_{1} = \frac{\lambda}{48}\,I_{2}(\sqrt\lambda) = e^{\sqrt\lambda}\bigg(\frac{\lambda^{3/4}}{48\sqrt{2\pi}}-\frac{5\lambda^{1/4}}{128\sqrt{2\pi}}+\cdots\bigg)
\ee
and we reproduce this expansion by
%\ba
%\vev{\mc W}_{1} &= \int_{-\infty}^{\infty}dt\,\frac{dz}{dt}\,g(z(t))\,e^{\frac{\sqrt\lambda}{2}[z(t)+\frac{1}{z(t)}]} \lp
%= \int_{-\infty}^{\infty}dt\,(i e^{it})\,\frac{-i}{32\pi}\bigg(\frac{1}{t^{4}}-\frac{i}{t^{3}}+\frac{1}{6t^{2}}+\cdots\bigg)\,
%e^{\sqrt\lambda (1-\frac{t^{2}}{2}+\frac{t^{4}}{24}\cdots)}  \lp
%%= \frac{e^{\sqrt\lambda}}{32\pi}\int_{-\infty}^{\infty}dt\,\bigg(\frac{1}{t^{4}}+\frac{2}{3}\frac{1}{t^{2}}+\cdots\bigg)\,e^{-\frac{\sqrt\lambda}{2} t^{2}+\frac{\sqrt\lambda}{24}t^{4}+\cdots} 
%%= \frac{\lambda^{-\frac{1}{4}}\,e^{\sqrt\lambda}}{32\pi}\int_{-\infty}^{\infty}dt\,\bigg(\frac{\lambda}{t^{4}}+\frac{2}{3}\frac{\sqrt\lambda}{t^{2}}+\cdots\bigg)\,
%%e^{-\frac{1}{2} t^{2}+\frac{1}{24\sqrt\lambda}t^{4}+\cdots} \lp
%=  \frac{\lambda^{-\frac{1}{4}}\,e^{\sqrt\lambda}}{32\pi}\int_{-\infty}^{\infty}dt\,\bigg[\frac{\lambda}{t^{4}}+\bigg(\frac{2}{3}\frac{1}{t^{2}}+\frac{1}{24}\bigg)\,\sqrt\lambda+\cdots\bigg]\,
%e^{-\frac{1}{2} t^{2}} \lp
%= \text{using (\ref{3.7}) for the pole terms and integrating directly the finite term }  \lp
%= \frac{\lambda^{-\frac{1}{4}}\,e^{\sqrt\lambda}}{32\pi}\,\bigg[
%\pi\frac{\lambda}{\Gamma(2+\frac{1}{2})}\left(\frac{1}{2}\right)^{2-1/2}+\bigg[-\pi\frac{2}{3}\frac{1}{\Gamma(1+\frac{1}{2})}
%\left(\frac{1}{2}\right)^{1-1/2}+\frac{1}{12}\sqrt\frac{\pi}{2}\bigg]\,\sqrt\lambda+\cdots\bigg] \lp 
%= \frac{\lambda^{3/4}}{48\sqrt{2\pi}}-\frac{5\lambda^{1/4}}{128\sqrt{2\pi}}+\cdots,
%\ea
%which is correct. An alternative method exploits 
the convenient change of parametrization 
\be
\la{3.12}
z+\frac{1}{z} = 2-u^{2}\ .
\ee
Using again $z=e^{it}$, this gives $u=2\sin\frac{t}{2}$ and one gets
\ba
\la{3.13}
\vev{\mc W}_{1} &= \frac{e^{\sqrt\lambda}}{\pi}
\int_{-\infty}^{\infty}du\,\frac{1}{u^{4}(4-u^{2})^{5/2}} e^{-\frac{\sqrt\lambda}{2}u^{2}}
= \frac{e^{\sqrt\lambda}}{\pi}\int_{-\infty}^{\infty}du\,\bigg(\frac{1}{32u^{4}}+\frac{5}{256u^{2}}+\cdots\bigg) e^{-\frac{\sqrt\lambda}{2}u^{2}}\\
&= \frac{e^{\sqrt\lambda}}{\pi}\,\pi\,\bigg[
\frac{1}{32}\frac{1}{\Gamma(2+\frac{1}{2})}\left(\frac{\sqrt\lambda}{2}\right)^{2-1/2}-\frac{5}{256}\,
\frac{1}{\Gamma(1+\frac{1}{2})}\left(\frac{\sqrt\lambda}{2}\right)^{1-1/2}+\cdots\bigg] \lp
=  \frac{\lambda^{3/4}}{48\sqrt{2\pi}}-\frac{5\lambda^{1/4}}{128\sqrt{2\pi}}+\cdots, \notag
\ea
in agreement with (\ref{3.11}).

\paragraph{Remark:} The integrals in (\ref{3.13}) are apparently divergent, even in Cauchy prescription. Actually, they are evaluated by formulas 
as (\ref{3.7}) that hide their original definition as finite contour integrals.

\section{Topological recursion for dominant strong coupling poles}
\la{sec:topological_recursion_strong}

We now look for a simplification of topological recursion (\ref{2.8}) based on considering the principal part of  resolvents at  $z_{i}=1$, \ie 
the terms that dominate at strong coupling.
Let us denote the highest pole part by $\hat\omega_{n,g}$. Introducing $\Delta_{i}=z_{i}-1$, the resolvents in  (\ref{2.11}) reduce to the compact expressions
\ba
\la{4.1}
\hat\omega_{1,1}(\Delta) &= \frac{1}{16\,\Delta^{4}},\notag \\
\hat\omega_{3,0}(\Delta_{1}, \Delta_{2}, \Delta_{3}) &= -\frac{1}{2\,\Delta_{1}^{2}\,\Delta_{2}^{2}\,\Delta_{3}^{2}},\notag \\
\hat\omega_{2,1}(\Delta_{1}, \Delta_{2}) &= \frac{5 \Delta _1^4+3 \Delta _1^2 \Delta _2^2+5 \Delta _2^4}{32 \Delta _1^6 \Delta _2^6}, \notag \\
\hat\omega_{1,2}(\Delta) &=-\frac{105}{1024\,\Delta^{10}}.
\ea
The (total) degree of the pole terms is $6(g-1)+4n$. In general, only even powers of $\Delta_{i}$ appear. If such an Ansatz is plugged into the topological recursion, one can compute the associated 
resolvent and project onto the maximal pole part. For instance, the last four resolvents in (\ref{2.10}) become, after projection,  
\ba
\la{4.2}
\hat\omega_{4,0}(\Delta_{1}, \Delta_{2}, \Delta_{3},  \Delta_{4}) &= -\frac{3 (\Delta _1^2 \Delta _2^2 \Delta _3^2+\Delta _1^2 \Delta _2^2
\Delta _4^2+\Delta _1^2 \Delta _3^2 \Delta _4^2+\Delta _2^2 \Delta 
_3^2 \Delta _4^2)}{4 \Delta _1^4 \Delta _2^4 \Delta _3^4 \Delta _4^4}, \notag \\
%%%%%%%%%%%%%%%%%%%%%%%%%%%%%%%%%%%%%%%%%%%%%%%%%%%%%%%%%%%%
\hat\omega_{3,1}(\Delta_{1}, \Delta_{2}, \Delta_{3}) &=  \frac{1}{64 \Delta _1^8 \Delta _2^8 \Delta _3^8}\,\bigg[35 \Delta _2^6 \Delta _3^6+30 \Delta _1^2 \Delta _2^4 \Delta 
_3^4 (\Delta _2^2+\Delta _3^2)\lp 
+5 \Delta _1^6 (7 \Delta _2^6+6 \Delta 
_2^4 \Delta _3^2+6 \Delta _2^2 \Delta _3^4+7 \Delta _3^6)+6 \Delta 
_1^4 (5 \Delta _2^6 \Delta _3^2+3 \Delta _2^4 \Delta _3^4+5 \Delta 
_2^2 \Delta _3^6)\bigg], \notag \\
%%%%%%%%%%%%%%%%%%%%%%%%%%%%%%%%%%%%%%%%%%%%%%%%%%%%%%%%%%%%
\hat\omega_{2,2}(\Delta_{1}, \Delta_{2}) &= -\frac{35 (33 \Delta _1^{10}+27 \Delta _1^8 \Delta _2^2+29 \Delta _1^6
\Delta _2^4+29 \Delta _1^4 \Delta _2^6+27 \Delta _1^2 \Delta _2^8+33 
\Delta _2^{10})}{2048 \Delta _1^{12} \Delta _2^{12}}, \notag \\
%%%%%%%%%%%%%%%%%%%%%%%%%%%%%%%%%%%%%%%%%%%%%%%%%%%%%%%%%%%%
\hat\omega_{1,3}(\Delta) &= \frac{25025}{32768 \Delta^{16}},
\ea
which are very compact expressions, compared with the full resolvents. Being symmetric functions, we can further simplify in terms of % Schur polynomials
elementary symmetric polynomials 
\be
\la{4.3}
e_{k}(x_{1}, \dots, x_{n}) = \sum_{1\le i_{1}<i_{2}<\cdots < i_{k}\le n}x_{i_{1}}\cdots x_{i_{k}},
\ee
where $x = \frac{1}{\Delta^{2}}$. One finds indeed the concise expressions
\ba
% \hat\omega_{4,0}&= 768\,s_{(2,1,1,1)}, \notag \\
\hat\omega_{4,0}&= -\frac{3}{4}  e_1 e_4~,\nonumber \\
%%%%%%%%%%%%%%%%%%%%%%%%%%%%%%%%%%%%%%%%%%%%%%%%%%%%%%%%%%%%
% \hat\omega_{3,1}&=  64\,(35\,s_{(4,1,1)}-5\,s_{(3,2,1)}-7\,s_{(2,2,2)}),  \notag \\
\hat\omega_{3,1}&=  \frac{35}{64} e_1^3 e_3 -\frac{75}{64} e_1 e_2 e_3 +\frac{33}{64} e_3^2 ~,\nonumber \\
%%%%%%%%%%%%%%%%%%%%%%%%%%%%%%%%%%%%%%%%%%%%%%%%%%%%%%%%%%%%
% \hat\omega_{2,2}&= 280\,(33\,s_{(6,1)}-6\,s_{(5,2)}+2\,s_{(4,3)}).
 \hat\omega_{2,2}&= -\frac{1155 }{2048}e_1^5 e_2 +\frac{2415}{1024}e_1^3 e_2^2-\frac{3955 }{2048} e_1 e_2^3 ~.
\ea
Further results are collected in Appendix \ref{app:resolvents}.

\paragraph{Remark:} Of course, the key point of the method is to use $\hat\omega$ projected resolvent in the topological recursion and never
using the full $\omega$'s.

\section{Large tension analysis of coincident Wilson loops}
\la{sec:coincW}

As a first application, we  consider the large tension limit of $\vev{\mc W^{n}}$ and, in particular,  the ratio (\ref{1.5}). As an illustration of the our strategy, we will begin with the doubly coincident Wilson loop, \ie the case $n=2$. 
Later, we shall extend the analysis to a generic number $n$ of coinciding loops.
For $n=2$, the $1/N$ expansion of $\vev{\mc W^{2}}$ has been considered in   \cite{Akemann:2001st,Plefka:2001bu,Okuyama:2018aij,Beccaria:2020ykg} and its first terms read
\ba
\frac{1}{N^{2}} & \vev{\mc W^{2}} = \left(\frac{2}{\sqrt\lambda}\,I_{1}\right)^{2}+\frac{\sqrt\lambda}{2N^{2}}\bigg(I_{0}I_{1}+\frac{1}{6}I_{1}I_{2}\bigg)  \lp
+\frac{1}{N^{4}}\bigg[\frac{37\lambda^{2}}{2304}I_{0}^{2}
-\frac{\sqrt\lambda(24+131\lambda)}{2880}\,I_{0}I_{1}+\frac{192+332\lambda+185\lambda^{2}}{11520}\,I_{1}^{2}\bigg]+\cdots\ ,
\ea
where $I_{n}\equiv I_{n}(\sqrt\lambda)$. The associated connected correlator is
\be
\vev{\mc W^{2}}_{\rm c} = \sum_{g=0}^{\infty}\frac{1}{N^{2g}}\vev{\mc W^{2}}_{{\rm c}, g} = \frac{\sqrt\lambda}{2}I_{0}I_{1}+\frac{1}{N^{2}}\bigg[
\frac{\lambda^{2}}{64}\,I_0^{2}-\frac{\lambda^{3/2}}{24} \, I_0\,I_{1}
+\frac{\lambda  (4+3 \lambda ) }{192} I_1^2
\bigg]+\cdots.
\ee
Expanding at large $\lambda$ and keeping the leading contribution at each order in $1/N$ gives
\be
\la{5.3}
\vev{\mc W^{2}}_{{\rm c},0} = \frac{1}{4\pi}\,e^{2\sqrt\lambda}\bigg(1-\frac{1}{4}\frac{1}{\sqrt\lambda}+\cdots\bigg),\qquad \vev{\mc W^{2}}_{{\rm c}, 1} = \frac{\lambda^{3/2}}{64\pi}\,e^{2\sqrt\lambda}\bigg(
1-\frac{19}{12}\frac{1}{\sqrt\lambda}+\cdots\bigg).
\ee
Let us show how these contributions can be easily recovered from the ``maximal poles'' topological recursion. We start from the 2-point formula
\be
\vev{\mc W^{2}}_{{\rm c}, g}^{\lambda\gg 1} = \frac{1}{(2\pi i)^{2}}\oint \hat\omega_{2,g}(z_{1}, z_{2})\,e^{\frac{\sqrt\lambda}{2}(z_{1}+1/z_{1})}\,e^{\frac{\sqrt\lambda}{2}(z_{2}+1/z_{2})}.
\ee
The genus 0 contribution is special being related to the universal Bargmann kernel and having no poles at $z_{1,2}=1$. It is 
\ba
\vev{\mc W^{2}}_{{\rm c}, 0}^{} &= \oint\frac{dz_{1}}{2\pi i}\frac{dz_{2}}{2\pi i}\frac{1}{(z_{1}-z_{2})^{2}}\,e^{\frac{\sqrt\lambda}{2}(z_{1}+1/z_{1})}\,e^{\frac{\sqrt\lambda}{2}(z_{2}+1/z_{2})} \notag \\
&= \oint\frac{dz_{1}}{2\pi i}\frac{dz_{2}}{2\pi i}\sum_{n=1}^{\infty}n\,z^{-1-n}\,w^{n-1}\sum_{p=-\infty}^{\infty}z^{p}I_{p}(\sqrt\lambda)\sum_{q=-\infty}^{\infty}w^{q}I_{q}(\sqrt\lambda)\notag \\
&= \sum_{n=1}^{\infty}n\,I_{n}^{2} = \frac{\sqrt\lambda}{2}\sum_{n=1}^{\infty}(I_{n-1}-I_{n+1})I_{n} = \frac{\sqrt\lambda}{2}\,I_{0}\,I_{1},
\ea
where in the last line we used the basic recursion of (modified) Bessel functions and the fact that the infinite sum is telescoping.
Starting at genus 1 we can apply the formula (\ref{3.7}) for the factorized poles. For instance, the first correction is 
\ba
\la{5.6}
\vev{\mc W^{2}}_{{\rm c}, 1}^{\lambda\gg 1} &= \oint\frac{dz_{1}}{2\pi i}\frac{dz_{2}}{2\pi i}\frac{1}{32}\bigg[
\frac{5}{(z_{1}-1)^{2}(z_{2}-1)^{6}}+\frac{3}{(z_{1}-1)^{4}(z_{2}-1)^{4}}\lp
+\frac{5}{(z_{1}-1)^{6}(z_{2}-1)^{2}}
\bigg]\,e^{\frac{\sqrt\lambda}{2}(z_{1}+1/z_{1})}\,e^{\frac{\sqrt\lambda}{2}(z_{2}+1/z_{2})} \stackrel{(\ref{3.7})}{\to}  e^{2\sqrt\lambda}\bigg[\frac{5}{32}(h_{1}h_{3}+h_{3}h_{1})+\frac{3}{32}h_{2}^{2}\bigg],
\ea
where the numerical constants $h_{m}$ are 
\be
\la{5.7}
h_{m} = \frac{(-1)^{m}}{\sqrt{2\pi}}\,\frac{\lambda^\frac{2m-1}{4}}{(2m-1)!!}.
\ee
Replacing (\ref{5.7}) in  (\ref{5.6}) reproduces the leading term in the second expression in  (\ref{5.3}). 

\paragraph{Extension to $\vev{\mc W^{n}}$ and high order calculation}

Similarly to (\ref{5.6}), we can exploit the resolvents in (\ref{4.1}) and (\ref{4.2}) (together with other ones in Appendix \ref{app:toprec-sub}) to 
evaluate the saddle point integrals needed to compute $\vev{\mc W^{n}}$
at high order in the genus expansion. Remarkably, this can be done for a generic $n$. To this aim, 
we introduce the variable
\be
\xi = \frac{\lambda^{\frac{3}{2}}}{8 N^{2}} = \frac{\pi\gs^{2}}{T},
\ee
and the connected correlators  
\ba
\vev{\mc W^{2}}_{c} &= -\vev{\mc W}^{2}+\vev{\mc W^{2}}, \qquad 
\vev{\mc W^{3}}_{c} = 2 \vev{\mc W}^3-3 \vev{\mc W}\vev{\mc W^{2}}+\vev{\mc W^{3}}, \notag \\
\vev{\mc W^{4}}_{c} &= -6 \vev{\mc W}^4+12 \vev{\mc W}^2 \vev{\mc W^2}-3 \vev{\mc W^2}^2-4 \vev{\mc W} \vev{\mc W^{3}}+\vev{\mc W^{4}},\ \textit{etc.}\ .
\ea
Normalizing by  suitable powers of the simple Wilson loop, we obtain the following results valid up to order $\mc O(\xi^{8})$:
\ba
\la{5.10}
  \ratio{2} &\LT \xi +\frac{\xi ^2}{3}+\frac{\xi ^3}{15}+\frac{\xi ^4}{105}+\frac{\xi ^5}{945}+\frac{\xi ^6}{10395}+\frac{\xi ^7}{135135}+\frac{\xi ^8}{2027025}+\cdots\ , \nonumber \\
  \ratio{3} &\LT 4 \xi ^2+\frac{14 \xi ^3}{3}+\frac{16 \xi ^4}{5}+\frac{169 \xi ^5}{105}+\frac{3046 \xi ^6}{4725}+\frac{532 \xi ^7}{2475}+\frac{290492 \xi ^8}{4729725}+\cdots\ , \nonumber \\
  \ratio{4} &\LT 32 \xi ^3+84 \xi ^4+\frac{1936 \xi ^5}{15}+\frac{138722 \xi ^6}{945}+\frac{635672 \xi ^7}{4725}+\frac{1078888 \xi ^8}{10395}+\cdots\ , \nonumber \\
  \ratio{5} &\LT 400 \xi ^4+\frac{5776 \xi ^5}{3}+5420 \xi ^6+\frac{2145776 \xi ^7}{189}+\frac{7815796 \xi ^8}{405}+\cdots\ , \nonumber \\
  \ratio{6} &\LT 6912 \xi ^5+54080 \xi ^6+247504 \xi ^7+\frac{53289280 \xi ^8}{63}+\cdots\ , \nonumber \\
  \ratio{7} &\LT 153664 \xi ^6+1804128 \xi ^7+\frac{185939824 \xi ^8}{15}+\cdots\ , \nonumber \\
  \ratio{8} &\LT 4194304 \xi ^7+\frac{209380864 \xi ^8}{3}+\cdots\ , \nonumber \\
  \ratio{9} &\LT 136048896 \xi ^8+ \cdots\ .
\ea
From connected correlators we obtain correlators of $n$ coincident Wilson loops using the combinatorial formula
\ba
  \la{5.11}
  \frac{\vev{\mc W^{n}}}{\vev{\mc W}^{n}}  &= 1 + \sum_{k < n}\sum_{\pi \in P(k,n)}\frac{n!}{S(\pi)(n-k-\abs{\pi})!
  \prod_{p \in \pi}(p+1)!}\prod_{p\in \pi}\ratio{p+1},
\ea
where $P(k,n)$ is the set of integer partitions $\pi$ of $k$ satisfying $k+\abs{\pi} \le n$ where $\abs{\pi}$ is the number of elements of $\pi$, and 
$S(\pi)$ is the symmetry factor of partition $\pi$ given by products of $m!$ for each group of $m$ equal elements in $\pi$. This expression follows from the fact that 
$\vev{\mc W^{n}}$ can be written as a sum over just the partitions of $n$. Since we divide by $\vev{\mc W}^{n}$, all the parts of a given partition that are
$1$ disappear, leaving the partitions of $n$ with every part at least $2$. Such partitions can be seen to be in one to one correspondence with the partitions of
integers $k \le n$ such that $k + \abs{\pi} \le n$, giving the expression above.

Since on general grounds $\ratio{m}=\mc O(\xi^{m-1})$, to obtain the expansion of $\frac{\vev{\mc W^{n}}}{\vev{\mc W}^{n}}$ up to
$\xi^{m}$ we can restrict the sum in (\ref{5.11}) to $k < m$. Furthermore, taking into account that a part $p$ enters the above expression as
$\ratio{p+1}$, we need the terms corresponding to all the partitions of $m$ to obtain the result up to $\xi^{m}$. Let's
consider a couple of examples. For $m=1$ the only possible partition is $1$ and we obtain 
\ba
  \la{5.12}
  \frac{\vev{\mc W^{n}}}{\vev{\mc W}^{n}} &= 1 + \frac{n!}{(n-2)!2!}\ratio{2} + O(\xi^{2}) =  1 + \frac{n(n-1)}{2}\ratio{2} + O(\xi^{2}).
\ea
For $m=2$ we have three partitions, \ie $1$ and $2$ and $(1,1)$. The last one has a symmetry factor of two. So we obtain
\ba
  \la{5.13}
  \frac{\vev{\mc W^{n}}}{\vev{\mc W}^{n}} &= 1 + \frac{n!}{(n-2)!2!}\ratio{2} +  \frac{n!}{(n-3)!3!}\ratio{3}
                                                +\frac{n!}{(n-4)!2!2!2!}\left(\ratio{2}\right)^{2}+ O(\xi^{3})  \\
                                              &=  1 + \frac{n(n-1)}{2}\ratio{2} +  \frac{n(n-1)(n-2)}{6}\ratio{3}
                                                +\frac{n(n-1)(n-2)(n-3)}{8}\left(\ratio{2}\right)^{2}+ O(\xi^{3}). \notag
\ea
In a similar way, to obtain the result up to $\xi^{3}$ we will have to add all terms corresponding to the partitions of three to the above results and so on.
We now have all the ingredients needed to evaluate the above expression to $\xi^{8}$. The final result is, \cf (\ref{1.5})
{\small
\ba
\la{5.14}
%  \frac{\vev{\mc W^{n}}}{\vev{\mc W}^{n}} &\LT
   & \mathsf{W}_{n}(\xi) = 1+ \frac{1}{2} (n-1) n\, \xi +\frac{1}{24} n \left(3 n^3-2 n^2-11 n+10\right) \xi ^2 \nonumber\\
                                              &+\frac{n}{720}\left(15 n^5+15 n^4-105 n^3-535 n^2+1674 n-1064\right) \xi ^3 \nonumber \\
                                              &+\frac{n}{40320} \left(105 n^7+420 n^6-70 n^5-13440 n^4-44303 n^3+401772 n^2-731028 n+386544\right) \xi ^4 \nonumber \\
                                              &+\frac{n}{241920} \big(63 n^9+525 n^8+1890 n^7-10710 n^6-177401 n^5-169715 n^4+8836872 n^3 \nonumber\\
                                              &+33525316 n^2+47031760 n-21987968\big) \xi ^5 \nonumber \\
                                              &+\frac{n}{159667200} \big(3465 n^{11}+48510 n^{10}+363825 n^9+596750 n^8-23242065 n^7-242099550 n^6+717147915 n^5
                                                \nonumber \\
                                              &+18424615770 n^4-131848499156 n^3+354391190648 n^2-421025682592 n+179605556480\big) \xi ^6 \nonumber \\
                                              &+\frac{n}{4151347200} \big(6435 n^{13}+135135 n^{12}+1606605 n^{11}+10385375 n^{10}-33778745 n^9-1632646015 n^8 \nonumber \\
                                              &-11172952505 n^7+124368186085 n^6+1239293411642 n^5-17142200059556 n^4+77299998069320 n^3 \nonumber \\
                                              & -168644525652096 n^2+177551331490176 n-70415438201856\big) \xi ^7 \nonumber \\
                                              &+\frac{n}{1394852659200} \big(135135 n^{15}+3963960 n^{14}+66846780 n^{13}+734894160 n^{12}+3276739466 n^{11} \nonumber \\
                                              &-61066061056 n^{10}-1348672905580 n^9-4090445158800 n^8+176081159789535 n^7+702493835315272 n^6 \nonumber \\
                                              &-24840194890564872 n^5+176744402329465152 n^4-616283230677646864 n^3+1159320084247595136 n^2 \nonumber \\
                                              &-1109932678089579264 n+414118538187171840\big) \xi ^8+ \cdots,
\ea
}
where the large tension limit is understood. This is the extension to order $\xi^{8}$ of the cubic result in Eq.~(1.17) of \cite{Beccaria:2020ykg}. The special cases $n=2$ and $n=3$ are 
\ba
\mathsf{W}_{2}(\xi) &= 1+\xi +\frac{\xi ^2}{3}+\frac{\xi ^3}{15}+\frac{\xi 
^4}{105}+\frac{\xi ^5}{945}+\frac{\xi ^6}{10395}+\frac{\xi 
^7}{135135}+\frac{\xi ^8}{2027025}+\cdots, \notag \\
\mathsf{W}_{3}(\xi) &=1+3 \xi +5 \xi ^2+\frac{73 \xi ^3}{15}+\frac{113 \xi 
^4}{35}+\frac{508 \xi ^5}{315}+\frac{33521 \xi ^6}{51975}+\frac{16139 
\xi ^7}{75075}+\frac{8803 \xi ^8}{143325}+\cdots, 
\ea
and agree with the exact expressions  \cite{Beccaria:2020ykg},
\ba
\la{5.16}
 \mathsf{W}_{2}(\xi) &= 1+e^{\frac{\xi}{2}}\sqrt\frac{\pi\,\xi}{2}\,\text{erf}\Big(
\sqrt\frac{\xi}{2}\Big), \notag \\
\mathsf{W}_{3}(\xi) &= 1+3e^{\frac{\xi}{2}}\sqrt\frac{\pi\,\xi}{2}\,\text{erf}\Big(
\sqrt\frac{\xi}{2}\Big)
+\frac{4\pi}{3\sqrt 3}\,\xi\,e^{2\xi}\,\Big[1-12\,T\Big(\sqrt{3\xi}, \frac{1}{\sqrt 3}\Big)\Big],
\ea
where 
\be
T(h,a)=\frac{1}{2\pi}\,\int_{0}^{a}\frac{ dx}{1+x^{2}} e^{-\frac{h^{2}}{2}\,(1+ x^2) },
\ee
 is the Owen T-function. The coefficient of $\xi^{k}$ is a polynomial in $n$ of degree $2k$. A remarkable simplification is achieved by writing (\ref{5.14})
 in exponential form 
 \ba
 \la{5.18}
 \log \mathsf{W}_{n}(\xi) &= \sum_{k=1}^{\infty}n\,(n-1)\, P_{k}(n)\,\xi^{k},
 \ea
 since $P_{k}$ turns out to be  a polynomial of (approximately half) degree $k-1$. Explicitly, one finds 
 \ba
 \la{5.19}
 P_{1} &= \frac{1}{2},\qquad
 P_{2} = -\frac{5}{12}+\frac{n}{6},\qquad 
 P_{3} = \frac{133}{90}-\frac{19 \,n}{18}+
\frac{n^2}{6},\notag \\
P_{4} &= -\frac{8053}{840}+\frac{3373 \,n}{360}-\frac{67 \,n^2}{24}+
\frac{n^3}{4},\notag \\
P_{5} &= \frac{171781}{1890}-\frac{34313 \,n}{315}+\frac{821 
\,n^2}{18}-\frac{118 \,n^3}{15}+\frac{7 \,n^4}{15}, \notag \\
P_{6} &= -\frac{20045263}{17820}+
\frac{7383058 \,n}{4725}-\frac{773086 \,n^2}{945}+\frac{6047 
\,n^3}{30}-\frac{419 
\,n^4}{18}+n^5,\notag \\
P_{7} &= \frac{694596731}{40950}-\frac{1177098997 
\,n}{44550}+\frac{230808223 \,n^2}{14175}-\frac{4806629 
\,n^3}{945}+\frac{534613 \,n^4}{630}-\frac{3001 \,n^5}{42}+\frac{33 
\,n^6}{14},\notag \\
P_{8} &= -\frac{45388623451}{152880}+\frac{2353546004957 \,n}{4633200}-
\frac{177458006053 \,n^2}{498960}+\frac{2001333893 
\,n^3}{15120}-\frac{183361073 \,n^4}{6480} \lp 
+\frac{2499191 
\,n^5}{720}-\frac{3609 \,n^6}{16}+\frac{143 \,n^7}{24},
  \ea
with  leading terms at large $n$ following  the pattern 
\be
P_{k} = \frac{4^{k-1}}{k\sqrt\pi}\frac{\Gamma(k-\frac{1}{2})}{\Gamma(k+2)}\,n^{k-1}+
\frac{4^{k-2}}{3k}\bigg[1-\frac{6(k-1)(2k-3)\Gamma(k-\frac{1}{2})}{\sqrt\pi\Gamma(k+2)}\bigg]\,n^{k-2}+
\cdots.
\ee

\subsection{Solution by Toda recursion}
\la{sec:toda}

The genus expansion of (\ref{1.3}) is efficiently computed by exploiting the Toda integrability of the
1-matrix Hermitian Gaussian model \cite{Morozov:2009uy}. In general, correlators in this model are constrained by integrable differential equations \cite{Morozov:1994hh,Morozov:1995pb,Mironov:2005qn}  
that in Gaussian case take the Toda form \cite{Gerasimov:1990is}. Notice that in \cite{Morozov:2009uy} the matrix model measure is 
 $\exp(-\frac{1}{2}\tr \widetilde M^{2})$ without explicit $N$ factor. This will be the convention throughout this section. After defining the connected correlators
\be
e(x_{1}, \dots, x_{k}) = \vev{\tr e^{x_{1}\widetilde M}\cdots \tr e^{x_{k}\widetilde M}}_{c},
\ee
one has 
\ba
& e_{N+1}(x)+e_{N-1}(x) = 2\,e_{N}(x) +\frac{x^{2}}{N}e_{N}(x), \\
\la{6.3}
& e_{N+1}(x,y)+e_{N-1}(x,y) = 2\,e_{N}(x,y) +\frac{(x+y)^{2}}{N}e_{N}(x,y)
-\frac{x^{2}y^{2}}{N^{2}}e_{N}(x) e_{N}(y).
\ea
The general structure is 
\be
e_{N+1}(x_{1}, \dots, x_{k})+e_{N-1}(x_{1}, \dots, x_{k})-\bigg[2+\frac{1}{N}(x_{1}+\cdots+x_{k})^{2}\bigg]\,e_{N}(x_{1}, \dots, x_{k})=g_{N}(x_{1}, \dots, x_{k}),
\ee
where $g_{N}$ may be read from the non-leading terms of the cumulant expansion of $\vev{X_{1}\cdots X_{k}}_{c}$ and replacing 
\be
\vev{X_{I_{1}}\cdots X_{I_{p}}}\to \frac{(x_{I_{1}}+\cdots+x_{I_{p}})^{2}}{N}\,e_{N}(x_{I_{1}}, \dots, x_{I_{p}}).
\ee
Here we are interested in the specialization to $x_{i} = \sqrt\frac{\lambda}{4N}$. Hence, defining
\be
e^{(k)}(\lambda, N) = e_{N}\left(\underbrace{\sqrt\frac{\lambda}{4N}, \dots, \sqrt\frac{\lambda}{4N}}_{k\ \text{terms}}\right),
\ee
we have the equations
\be
e^{(k)}\left(\frac{N+1}{N}\,\lambda, N+1\right)+e^{(k)}\left(\frac{N-1}{N}\,\lambda, N-1\right)-\bigg(2+\frac{k^{2}\lambda}{4N^{2}}\bigg)\,e^{(k)}(\lambda, N)=g^{(k)}(\lambda, N),
\ee
where $g^{(k)}(\lambda, N)$ is obtained from the non-leading terms of the cumulant expansion of $\vev{X^{k}}_{c}$ and replacing 
\be
\vev{X^{p}}\to \frac{p^{2}\lambda}{4N^{2}}\,e^{(p)}(\lambda, N).
\ee
The explicit coefficients of the cumulant expansion of $\vev{X^{p}}_{c}$ may be expressed in terms of integer partitions $\pi=(1^{m_{1}}2^{m_{2}}\cdots)$ of $p$
\be
\vev{X^{p}}_{c} = \sum_{\pi\in\mc P(p)}(-1)^{|\pi|-1}(|\pi|-1)! \sigma(\pi)\, \prod_{r}\vev{X^{r}}^{m_{r}}, \qquad \sigma(\pi) = \frac{p!}{\prod_{r}(r!)^{m_{r}} m_{r}!}.
\ee
Hence, the equations are 
\ba
& e^{(k)}\left(\frac{N+1}{N}\,\lambda, N+1\right)+e^{(k)}\left(\frac{N-1}{N}\,\lambda, N-1\right)-\bigg(2+\frac{k^{2}\lambda}{4N^{2}}\bigg)\,e^{(k)}(\lambda, N)\notag \\
&=\mathop{\sum_{\pi\in\mc P(p)}}_{ \pi\neq (p)}(-1)^{|\pi|-1}(|\pi|-1)! \sigma(\pi)\prod_{r}\left(\frac{r^{2}\lambda}{4N^{2}}\,e^{(r)}(\lambda, N)\right)^{m_{r}}.
\ea
The large tension scaling Ansatz is 
\be
e^{(k)}(\lambda, N) = e^{k\sqrt\lambda}F_{k}(\xi), \qquad \xi = \frac{\lambda^{3/2}}{8N^{2}}.
\ee
Replacing in the Toda equations gives
\ba
&  e^{k\sqrt\lambda}\bigg[\frac{k}{12}\xi^{1/3}(k^{3}\xi-6)F_{k}(\xi)-k\,\xi^{4/3}\,F_{k}'(\xi)\bigg]N^{-4/3}+\cdots \lp
=\mathop{\sum_{\pi\in\mc P(p)}}_{ \pi\neq (p)}(-1)^{|\pi|-1}(|\pi|-1)! \sigma(\pi)\prod_{r}\left(r^{2}\xi^{2/3}N^{-2/3}e^{r\sqrt\lambda}\,F_{r}(\xi)\right)^{m_{r}}.
\ea
The only partitions that may give a contribution have $|\pi|=\sum m_{r}=2$. One case is when $k$ is even and then the partition is $\pi = (\frac{M}{2}, \frac{M}{2})$, or when $k$ is split
into the sum of two different parts $\pi = (q, (k-q))$ with $q\neq k/2$. \footnote{Notice that distinct partitions are ordered.}  
Denoting by an apex such partitions, we have (using $\sum_{r}r m_{r}=k$)
\ba
& \frac{k}{12\xi}(k^{3}\xi-6)F_{k}(\xi)-k\,F_{k}'(\xi)
=\mathop{\sum'_{\pi\in\mc P(p)}}_{ \pi\neq (p)}(-1)^{|\pi|-1}(|\pi|-1)! \sigma(\pi)\prod_{r}\left(r^{2}\,F_{r}(\xi)\right)^{m_{r}}.
\ea
Finally, evaluating the r.h.s. for the two relevant kinds of partitions, we obtain the differential equation
\ba
& \frac{k}{12\xi}(k^{3}\xi-6)F_{k}(\xi)-k\,F_{k}'(\xi)
=-\frac{1}{2}\sum_{p=1}^{k-1}\binom{k}{p}\,p^{2}(k-p)^{2}F_{p}(\xi)F_{k-p}(\xi)\ ,
\ea
that we rearrange in the form
\ba
& F_{k}'(\xi)+\frac{1}{12\xi}(6-k^{3}\xi)F_{k}(\xi)
=\frac{1}{2k}\sum_{p=1}^{k-1} \binom{k}{p}\,p^{2}(k-p)^{2}F_{p}(\xi)F_{k-p}(\xi).
\ea
The first instance $k=1$ gives 
\be
F_{1}'(\xi) +\frac{1}{12\xi}(6-\xi) F_{1}(\xi) = 0 \quad \to \quad F_{1}(\xi) = C\,\xi^{-1/2}e^{\frac{\xi}{12}}.
\ee
The constant is fixed by (\ref{1.3}) and gives
\be
F_{1}(\xi) = \frac{1}{2\sqrt\pi}\,\xi^{-1/2}e^{\frac{\xi}{12}}.
\ee
We shall be interested in the ratios 
\be
R_{k}(\xi) = \frac{F_{k}(\xi)}{F_{1}(\xi)^{k}}.
\ee
They obey 
\ba
& R_{k}'(\xi)-\frac{k-1}{12\xi}\bigg[6+k\,(k+1)\,\xi\bigg]\,R_{k}(\xi)
=\frac{1}{2k}\sum_{p=1}^{k-1}\binom{k}{p}\,p^{2}(k-p)^{2}R_{p}(\xi)R_{k-p}(\xi).
\ea
We also know that $R_{k}(\xi) = \mc O(\xi^{k-1})$. This gives the integration constant and the explicit recurrence relation
\ba
\la{6.20}
R_{1}(\xi) &= 1, \notag \\
R_{k}(\xi) &= e^{\frac{k(k^{2}-1)}{12}\,\xi}\,\xi^{\frac{k-1}{2}}\,\int_{0}^{\xi}dz\,e^{-\frac{k(k^{2}-1)}{12}z}\,z^{\frac{1-k}{2}}
\frac{1}{2k}\sum_{p=1}^{k-1}\binom{k}{p}\,p^{2}(k-p)^{2}R_{p}(z)R_{k-p}(z).
\ea
This recursion provides the expressions in (\ref{5.10}) to be plugged into (\ref{5.11}) in order to compute the scaling functions $\mathsf{W}_{n}(\xi)$.
Just to give an example, using (\ref{6.20}) one may easily extend the last line in (\ref{5.10}) and find 
{\small
\ba
\ratio{9} &\LT 136048896\, \xi ^8+3073838592\, \xi ^9+\frac{203451048576}{5}\,\xi^{10}+\frac{14167449806208}{35}\, \xi ^{11}\\
& +\frac{579258140455408 }{175}\,\xi ^{12}+\frac{26870941356966016 }{1155}\,\xi^{13}+\frac{487595584858030932224}{3378375}\, \xi^{14}\lp
+\frac{1672326501977995232 }{2079}\xi^{15}+\frac{140563211710013333686888 }{34459425}\xi^{16}
+\frac{24326465124436842115824 }{1280125}\,\xi^{17}\lp
+\frac{13943786316178585925001407879336}{170147718365625}\, \xi^{18}
+\frac{22599997060403576152832986605968 }{68656096884375}\xi^{19}+\cdots.\notag
\ea
}
This allows to compute the polynomials in (\ref{5.19}) at higher order. For instance
\ba
\la{uu1}
P_{9} &= \frac{4199422654038881}{723647925}-\frac{1832589301073441 
\,n}{170270100}+\frac{3565245577877609 
\,n^2}{425675250}-\frac{372258099643 \,n^3}{103950}\lp 
+\frac{26122674017 
\,n^4}{28350}-\frac{828493349 \,n^5}{5670}+\frac{3760909 
\,n^6}{270}-\frac{19652 \,n^7}{27}+\frac{143 \,n^8}{9}, \notag \\
%%%
P_{10} &= -\frac{3360311016680854273}{27498621150}+\frac{21169992523320924583 
\,n}{86837751000}-\frac{118100410140883837 
\,n^2}{567567000} \lp
+\frac{85134261154188137 
\,n^3}{851350500}-\frac{27943401961331 \,n^4}{935550}+\frac{81563996453 
\,n^5}{14175}-\frac{225150754 \,n^6}{315}\lp  
+\frac{4961927 
\,n^7}{90}-\frac{14341 \,n^8}{6}+\frac{221 \,n^9}{5} \ ,
\ea
and so on. Further expressions of $P_{k}$ for $k$ up to 20 are collected in Appendix \ref{app:ppp}. 

\paragraph{Remark:} Of course, one can also use (\ref{6.20}) without expanding. This gives exact expressions for $R_{k}$ as iterated integrals. The first two cases are
\ba
R_{2}(\xi) &= \sqrt\frac{\pi\xi}{2}\,e^{\xi/2}\,\text{erf}\left(\sqrt\frac{\xi}{2}\right), \notag \\
R_{3}(\xi) &= -\frac{4\pi}{3\sqrt 3}\xi\,e^{2\xi}\,\bigg[-1+12\,T\left(\sqrt{3\xi}, \frac{1}{\sqrt 3}\right)\bigg],
\ea
where $T$ is the Owen function, \cf (\ref{5.16}). The expression for $R_{4}$ may be obtained by continuing the iteration but will involve integrals
of the $T$ function. A simple general feature of the functions $R_{k}$ is that they are all entire in $\xi$. Hence, the radius of convergence of (\ref{5.14})
is infinite for all $n$.

\paragraph{Remark:} One has to keep in mind that Toda recursion methods are not suitable to treat insertions of local chiral operators, see the discussion in 
Appendix \ref{app:toda-chiral}. 
In this case,  one has to keep using topological recursion, as discussed in the next Section.

\section{Correlator of coincident Wilson loops and a   chiral operator}
\la{sec:chiral}

In this section, we address the problem of computing the correlator between multiple coincident Wilson loops $\mc W^{n}$
and a single trace chiral operator. In other words, we want to generalize (\ref{1.7}) and prove (\ref{1.11}), (\ref{1.12}).
To properly define chiral primaries let us recall that the  $\frac{1}{2}$-BPS Wilson loop, associated
with $\tr\big(e^{\frac{\sqrt\lambda}{2} M}\big)$ in the Gaussian matrix model, \cf (\ref{1.1}), stands for the operator
\be
\mc W = \tr \mc P \exp\bigg\{g_{\rm YM}\int_{\mc C} d\sigma\,[i\,A_{\mu}(x)\,\dot x^{\mu}(\sigma)+\, R\,\Phi_{1}(x)]\bigg\},
\ee
where $\mc C$ is a circle of radius $R$ (set to unity in the following), and $\Phi_{1}$ is one of the six real scalars $\{\Phi_{I}\}_{I=1, \dots, 6}$ in $\mc N=4$ SYM.
Single trace chiral operators take the general form $\mc O_{J} = \tr(u_{I}\Phi_{I}(x))^{J}$ where 
$u_{I}$ is a complex null 6-vector obeying $u^{2}=0$ \cite{Semenoff:2001xp}. The dependence of the correlator $\vev{\mc W\,\mc O_{J}}$ on $u_{I}$ and the 
choice of  coupling between the loop and the scalars factorizes and will be absorbed in the operator normalization \cite{Maldacena:1998im}. 
With the same conventions as in \cite{Beccaria:2020ykg}, the matrix model representative for the chiral operator $\mc O_{J}$ is
\be
\la{7.2}
\mc O_{J} = \frac{N}{2}\,\bigg(\frac{\pi}{2}\bigg)^{J/2-1}:\tr M^{J}:  \ , 
\ee
where normal ordering subtracts self-contractions and is necessary to map matrix model correlators to $\mathbb R^{4}$ quantum
expectation values \cite{Billo:2019job,Billo:2017glv}. \footnote{The choice of normalization in (\ref{7.2}), and in particular the overall power of $N$, 
is dictated by string theory and makes direct contact with the associated natural vertex operators \cite{Giombi:2020mhz}. Another 
standard choice is to require a fixed normalization of the chiral operators 2-point functions as in \cite{Semenoff:2001xp}.
}
At leading order in large tension, the correlator between a single Wilson loop and the chiral operator $\mc O_{J}$ obeys (\ref{1.7}) in terms of a scaling function that depends on 
the specific ratio $\gs^{2}/T^{2}$ and has a non-trivial dependence on $J$. The most natural scaling dependence is actually on $\gs^{2}/T$ as in (\ref{1.5}).
Several cancellations occur and are responsible for the relevant variable being $\gs^{2}/T^{2}$. We shall show that this pattern 
changes in the case of  the correlator between multiple coincident Wilson loops and one chiral operator. The above mentioned  cancellations do not occur anymore
and one has instead the structure (\ref{1.11}). Besides, 
the function $\mathsf{H}_{n}$ can be computed explicitly in terms of $\mathsf{W}_{n}$, \cf (\ref{1.12}).
To derive such a result, we will conveniently use the strong coupling version of topological recursion. As we remarked previously,
Toda recursion is rather cumbersome for these purposes, as illustrated in the example 
$(n,J) = (1,2)$ in Appendix~\ref{app:toda-chiral}.

\subsection{Contribution from multi-trace operators in normal ordering}

As a preliminary step we first address the issue of the effects of normal ordering in (\ref{7.2})
and the role of multi-trace operators. It is instructive to look at the first cases at low  $J$. A straightforward explicit calculation 
gives (we restrict to even $J$ for the purpose of illustration)
\ba
:\tr M^{2}: &= \tr M^{2} -\frac{N}{2}, \qquad :\tr M^{4}: = \tr M^{4} -2\,\tr M^{2}-\frac{1}{N}\,(\tr M)^{2}+\frac{N}{2}+\frac{1}{4N}, \notag \\
:\tr M^{6}: &= \tr M^{6}-3\,\tr M^{4}+\bigg(\frac{15}{4}+\frac{15}{4N^{2}}\bigg)\,\tr M^{2}  \lp
+\frac{1}{N}\bigg(-\frac{3}{2}\,(\tr M^{2})^{2}-3\,\tr M\,\tr M^{3}+\frac{15}{4}\,(\tr M)^{2}
\bigg) -\frac{5N}{8}-\frac{5}{4N},
\ea
and so on. In general, terms involving products of $k$ traces are $\sim 1/N^{k-1}$ at large $N$. 
We will write  
\ba
\la{7.4}
:\tr M^{J}:\,\,\, = \sum_{k \ge 0}\frac{1}{N^{k-1}}\,\left[:\tr M^{J}:\right]_{k}~.
\ea
where the operators in $\left[:\tr M^{J}:\right]_{k}$ have coefficients $\mc O(1)$ at large $N$.
\footnote{The $k$-trace part may have an explicit $N$ dependence as in $\left[:\tr M^{6}:\right]_{1}$
which has a piece $\frac{15}{4}(1+1/N^{2})\tr M^{2}$ whose $N\to \infty$ limit is finite. }
Now, let us consider the genus $g$ contribution to the connected correlator 
 $\vev{\mathcal{W}^{n}\mathcal{O}^{(k)}}_{c}$ where $\mathcal{O}^{(k)}$ is any arbitrary $k$-trace operator.
 We can write, \cf (\ref{2.5}), 
 \ba
  \la{7.5}
&  \left. \vev{\mathcal{W}^{n}\mathcal{O}^{(k)}}_{c}\right|_{{\rm genus}\ g}  \\ 
&\qquad =  \frac{1}{N^{2g+k+n-2}(2\pi i)^{k+n}}\oint \, \omega_{n,g}\left(z_{1} ,\dots ,z_{n+k}\right)
                                                 \mathcal{O}^{(k)}(x(z_{1}), \dots , x(z_{k}))\exp\left(\frac{\lambda}{2}\sum_{l=1}^{n}x\left(z_{k+l}\right) \right)~.
                                                 \notag
\ea
In the case of $\mc O^{(k)} = [:\tr M^{J}:]_{k}$, taking into account the extra factor  $\frac{1}{N^{k-1}}$ in (\ref{7.4}),
we find that $\vev{\mathcal{W}^{n}[:\tr M^{j}:]_{k}}_{c}|_{{\rm genus}\ g}$ scales as $N^{-(2g + 2k + n -3)}$.
Finally, let us pin the dependence on $\lambda\gg 1$. Since the operator $:\tr M^{J}:$ does not depend on $\lambda$ explicitly, the  strong coupling limit of the expectation value
of Wilson loop with chiral operators corresponds  to maximizing
the order of poles of variables corresponding to Wilson loop or conversely minimizing the order of the poles of the variables that correspond to
the $:\tr M^{J}:$ operator.
%\footnote{The minimization comes with the constraint that the order can't be zero, since those terms don't contribute to expectation value}. 
The total order of the poles of $\omega_{n,g}$ is $6(g-1)+4n$, as discussed in section  \ref{sec:topological_recursion_strong}.
According to the saddle point analysis this implies the final scaling behaviour
\ba
  \la{7.6}
  \left. \vev{\mathcal{W}^{n}[:\tr M^{J}:]_{k}}_{c}\right|_{{\rm genus}\ g} \sim \frac{\lambda^{\frac{1}{4}(6g + 3n + 2k - 6)}}{N^{2g+2k+n-3}}.
\ea
This gives the \underline{leading} power at large $\lambda$ for all genera. In particular, a term with an overall $\frac{1}{N^{P}}$ factor will be accompanied 
by the following powers of $\lambda$
\be
\lambda^{\frac{3}{4}-k}\,\bigg(\frac{\lambda^{\frac{3}{4}}}{N}\bigg)^{P},
\ee
showing that multiple trace contributions are suppressed. Besides, since the saddle point expansion has relative corrections in powers $\lambda^{-1/2}\sim 1/T$, 
double trace corrections to normal ordering cannot be seen even at first subleading order in large $\lambda$.

Let us see this explicitly in the simplest case of a single Wilson loop keeping only up to double trace operators .
At the planar level, as is well known, the double trace part doesn't contribute. At  $1/N^{2}$ level there are two relevant cumulants corresponding to
$(k,g)=(1,1)$ and $(2,0)$.
 Their $\lambda$ dependence can be obtained using the explicit strong coupling resolvents given \eqref{4.2} as respectively,
\ba
  \la{7.8}
  \frac{1}{N^{2}}\oint_{z_{2}} \exp\left(\frac{\sqrt{\lambda}}{2}x(z_{2})\right)\hat{\omega}_{2,1}(z_{1},z_{2}) \sim \frac{\lambda^{\frac{5}{4}}}{N^{2}},%
  \qquad
    \frac{1}{N^{2}}\oint_{z_{3}} \exp\left(\frac{\sqrt{\lambda}}{2}x(z_{3})\right)\hat{\omega}_{3,0}(z_{1},z_{2},z_{3}) \sim \frac{\lambda^{\frac{1}{4}}}{N^{2}}.
\ea
The first contribution is dominant in the large tension limit and in fact we would have to expand it to three orders in $\lambda$ before the the
second one becomes effective. Since, the rate of growth of the exponent of $\lambda$ is $6$ for $g$ but only $2$ for $k$, as $g$ and $k$ increase or as cumulants are multiplied, the gap between the contributions of single and higher trace operators only increases. \footnote{This is
of course contingent on these first two contributions from single trace operators not vanishing once we take residue integrals corresponding to the chiral
operator. As we now show this is indeed the case.
This remark will be important later.}

\subsection{\texorpdfstring{$\vev{\mathcal{W}^{n}\mathcal{O}_{J}}$}{WOJ} at leading order}\la{sec:evwno-at-leading}
As a result of the above discussion, we can restrict ourselves to  
the single trace part of  normal ordering, \ie the planar approximation. 
According to \cite{Rodriguez-Gomez:2016cem,Beccaria:2020hgy}, it may be written in terms  
of  Chebyshev polynomials and, in the $z$ variable, it reads
\ba
  \la{7.9}
  :\tr M^{J}:\,\, \to z^{J} + \frac{1}{z^{J}} + 2 \delta_{j,2} + \dots ~.
\ea
The relevant connected correlators $\vev{\mathcal{W}^{n} :\tr M^{J}:}_{c,g}$ are \footnote{
We deform the $z_{1}$ integration shrinking it around $z_{1}=0$. This is possible because residues vanish at $z_{i}=\pm 1$ and, in particular,  
we can drop the parts that are not singular at $z_{1}=0$.}
\ba
  \la{7.10}
  \vev{\mathcal{W}^{n} \mathcal{O}}_{c,g} &=  \frac{1}{(2\pi i)^{n+1}}\oint \omega_{n,g}(z_{1} , \dots , z_{n+1})
  \,z_{1}^{-J}\,
  \exp\left[\frac{\sqrt{\lambda}}{2} (x(z_{2}) + \dots + x(z_{n+1}))\right] ~.
\ea
In the strong coupling limit we use the strong coupling resolvent $\hat{\omega}_{n,g}(z)$ and keep in it only those terms that minimize the order of
the poles of $z_{1}$ at $1$. This can be done by going through one step of topological recursion. This  corresponds to starting with
$\omega_{g,n-1}(z_{1} , \dots z_{n})$ and using:
\ba
  \la{7.11}
  \prod_{i=2}^{n+1}\frac{1}{(z_{i}-1)^{2 k_{i}}} \to \sum_{j=2}^{n}\frac{2k_{j}+1}{4(z_{1}-1)^{2}(z_{j}-1)^{2}}\prod_{i=2}^{n+1}\frac{1}{(z_{i}-1)^{2 k_{i}}}+ \cdots ~.
\ea
We can now integrate over $z_{2},\cdots,z_{n}$ in the saddle point approximation. The above factor of $2k_{j}+1$  ensures that the result has a very simple
relation to $\vev{\mathcal{W}^{n}}_{c,g}$ in the strong coupling limit, \ie \footnote{Let us remind that the presence of $dz_{1}$ in the residue
is formal at this level and could be omitted. Nevertheless, it is convenient to keep it to emphasize transformation properties under change of variables.}
\ba
  \la{7.12}
  \vev{\mathcal{W}^{n}:\tr M^{J}:}_{c,g}^{\lambda\gg 1}  &=
   \frac{n\lambda}{2}\vev{\mathcal{W}^{n}}^{\lambda \gg 1}_{c,g}
   \mathop{\Res}_{z_{1}=0}\frac{d z_{1}}{(z_{1}-1)^{2}z_{1}^{J}} + \dots
                                                    = \frac{J\, n\, \sqrt{\lambda}}{2}\vev{\mathcal{W}^{n}}^{\lambda \gg 1}_{c,g} + \cdots~.
\ea
The same is true for the full correlator, after expanding  into connected correlators, i.e.
\ba
  \la{7.13}
\frac{\vev{\mathcal{W}^{n}:\tr M^{J}:}}{\vev{\mathcal{W}^{n}}}  &\stackrel{\lambda\gg 1}{=} \frac{J\, n\, \sqrt{\lambda}}{2}+ \cdots ~.
\ea
This is of course expected from the known results for $n=1$ and $n=2$, see  \cite{Beccaria:2020ykg}.

\subsection{Subleading corrections}\la{sec:subl-corr-1}

%Naively, beyond the leading order result (\ref{7.13}), one expects a non-trivial series  in the ratio $\gs^{2}/T$. 
%by using just the strong coupling resolvents $\hat{\omega}_{n,g}$. 
%Actually, at subleading orders, cancellations occur and reorganize the r.h.s. of (\ref{7.13}) 
%in the form \eqref{1.7}, \ie a power series in $\gs^{2}/T^{2}$. 
%\footnote{
%Notice also that , as explained above, the only possible
%poles of $z-1$ that can contribute in this limit are of order $2$. Hence, the only possible dependence on $J$ that can be obtained in this limit is an overall
%factor of $J$ which goes against the results obtained for $n=1,2$ in \cite{Beccaria:2020ykg}. The fact that instead of a non-trivial series we only obtain a
%single term from \eqref{7.13} is the first sign of cancellations mentioned earlier in the introduction. 
%Remarkably, for $n>1$ this happens in a simple way as we are going to illustrate.

To go beyond leading order
we need to carry out topological recursion with poles of one subleading order included. It is convenient to change
variables  from $z$ to $u$, \cf  \eqref{3.12},  and write
\ba
  \la{7.14}
  \omega_{n,g}(u_{1} , \dots , u_{n}) &= \hat{\omega}_{n,g} + \delta \omega_{n,g} + \dots~.
\ea
Where $\delta \omega_{n,g}$ includes the poles of total degree $6g + 4n -8$, see Appendix \ref{app:toprec-sub}
for full details of the procedure. Using (\ref{7.14})  we
can compute the one-variable resolvents obtained after integration of  all but one variable, 
\footnote{We change back to $z$-coordinates for the free variable because 
this is convenient  to compute expectation values with a chiral operator.}
\ba
  \la{7.15}
  \bar{\omega}_{n,g}(z) &= \frac{1}{(2\pi i)^{n}} \oint \omega_{n+1,g}(u(z) , u_{1} , \dots u_{n})~.
\ea
Due to our previous discussion, \cf (\ref{7.6}), the first two orders in the $1/\sqrt{\lambda}$ expansion at large $\lambda$
can be computed by  ignoring mixing with multi-trace operators and using the simple correspondence 
in (\ref{7.9}). Thus, we simply obtain
\ba
  \la{7.16}
  \vev{\mathcal{W}^{n}:\tr M^{J}:}_{c,g} &= \mathop{\Res}_{z \to 0} \frac{\bar{\omega}_{n,g}(z)}{z^{J}} ~.
\ea
This computes the connected part of the correlator but we can also define a function that similarly computes the full  correlator, \ie
\ba
  \bar{\Omega}_{n,g}(z) = \sum_{k=1}^{n}\sum_{h=0}^{g} \binom{n}{k} \vev{\mathcal{W}^{n-k}}_{g-h}\bar{\omega}_{k,h}(z) ~\   \to  \ 
  \la{7.17}
  \vev{\mathcal{W}^{n}:\tr M^{J}:} &= \mathop{\Res}_{z\to 0}\sum_{g=0}^{\infty}\frac{1}{N^{2g-1}}\frac{\bar{\Omega}_{n,g}(z)}{z^{J}}~.
\ea
To compute  $\vev{\mathcal{W}^{n}:\tr M^{J}:}/\vev{\mathcal{W}^{n}}$ it is convenient to expand $\bar{\Omega}_{n,g}(z)$ as:
\ba
  \la{7.18}
  \bar{\Omega}_{n,g}(z) &= U_{n,0}(z)\vev{\mathcal{W}^{n}}_{c,g} + U_{n,1}(z) \vev{\mathcal{W}^{n}}_{c,g-1} 
 % + \dots + U_{n,h}(z) \vev{\mathcal{W}^{n}}_{c,g-h}
                               + \dots + U_{n,g}(z) \vev{\mathcal{W}^{n}}_{c,0} + \dots~,
\ea
where each $U_{n,g}(z)$ is determined recursively genus by genus and final dots stand 
for a correction  of order $\mc O((1/\sqrt\lambda)^{g+2})$ relative to the leading order. 
Then it can be seen that,
\ba
  \la{7.19}
  \frac{\vev{:\tr M^{J}:\mathcal{W}^{n}}}{\vev{\mathcal{W}^{n}}} &= \mathop{\Res}_{z\to 0}\sum_{g=0}^{\infty}\frac{1}{N^{2g-1}}\frac{U_{n,g}(z)}{z^{J}}~.
\ea
The functions $U_{n,g}(z)$ depend also on $\lambda$. To the leading order in $\lambda$, $U_{n,0}(z)$ can be read from \eqref{7.12}. To get a non-vanishing result for all other
$U_{n,g}$ we need to go beyond $\hat{\omega}_{n,g}$ and include 
$\delta \omega_{n,g}$. Restricting ourselves to two leading term in $\lambda$, the most general
structure possible for $U_{n,g}$ is:
\ba
  \la{7.20}
  U_{n,0}(z) &= d z \left(\frac{n \sqrt{\lambda} }{2(z-1)^{2}} + \frac{f_{n,0}(z)}{(z-1)^{4}}\right) + \dots ~,\nonumber \\
  U_{n,g}(z) &=  \frac{d z f_{n,g}(z) \lambda^{\frac{3g}{2}}}{(z-1)^{4}} + \dots~.
\ea
Where $f_{n,g}(z)$  are polynomials of degree at most $3$ and independent of $\lambda$. Two out of the 4 free coefficients are determined 
by the requirement from topological recursion that
$U_{n,g}\left(\frac{1}{z}\right) = - U_{n,g}(z)$. Another one can be fixed by requiring that $\vev{:\tr M: \mc W^{n}}_{c,g}$ vanishes
for $g > 0$. \footnote{This may be shown by explicit splitting of $U(N)$ into $U(1)\times SU(N)$, see Appendix \ref{app:sec:corr-funct-vev}.}. Combining these two requirements we obtain
\ba
  \la{7.21}
  U_{n,g} &= d z\,\frac{c_{n,g}\lambda^{\frac{3g}{2}} z}{(z-1)^{4}}~.
\ea

\paragraph{Explicit results}
\la{sec:results-wilson-chiral}

After having clarified the general structure of topological recursion for the quantities we need, let us 
present explicit results. For the 'critical' case $n=1$ we find \footnote{Recall that we expect a major change
of features when moving from $n=1$ to $n>1$.}
\ba
  \la{7.22}
  U_{1,0}(z) &= dz\, \left(\frac{\sqrt{\lambda }}{2  (z-1)^2}+\frac{3 z}{2  (z-1)^4}\right) + \dots\nonumber ~, \\
  U_{1,1}(z) &= dz\,\frac{\lambda ^{3/2} z}{32  (z-1)^4} + \dots~,
\ea
while the higher $U_{1,g}(z)$ vanish i.e $c_{1,g} = 0$ for $g > 1$. This can be seen as consistency check and is a result of the cancellations required to
reorganize the series for $\vev{\mathcal{W}:\mathcal{O}_{J}:}$ as in \eqref{1.7}. To calculate non-vanishing terms in $U_{1,g}(z)$ for $g > 1$ we will need to
keep more than $2$ leading terms in $\omega_{n,g}$. 

These peculiar cancellations do not occur for $n > 1$ and make the calculation of subleading corrections
possible with our level of accuracy. We find 
\ba
  \la{7.23}
  \sum_{g}\frac{1}{N^{2g-1}}U_{2,g}(z) &= \frac{1}{N}\bigg( \frac{\sqrt{\lambda }}{ (z-1)^2}-\frac{3 z}{ (z-1)^4}
                                      + \frac{z}{(z-1)^{4}} \Big[\frac{7 \lambda ^{3/2}}{16 N^2}-\frac{\lambda ^3}{64 N^4}+\frac{3 \lambda ^{9/2}}{2560 N^6}~,\nonumber \\
  &\phantom{abcdef}-\frac{37 \lambda ^6}{430080 N^8}+\frac{13 \lambda ^{15/2}}{2064384 N^{10}}-\frac{299 \lambda ^9}{648806400 N^{12}}\Big]\bigg)+ \dots \nonumber \\
  \sum_{g}\frac{1}{N^{2g-1}}U_{3,g}(z) &= \frac{3}{2 N}\bigg(\frac{ \sqrt{\lambda }}{ (z-1)^2}-\frac{3 z}{  (z-1)^4}
                                      + \frac{z}{(z-1)^{4}}\Big[\frac{13 \lambda ^{3/2}}{16 N^2}+\frac{\lambda ^3}{32 N^4}~,\nonumber \\
  &\phantom{abcdef}-\frac{17 \lambda ^{9/2}}{1280 N^6}+\frac{123 \lambda ^6}{71680 N^8}-\frac{\lambda ^{15/2}}{516096 N^{10}}\Big]\bigg) + \dots ~, \nonumber \\
  \sum_{g}\frac{1}{N^{2g-1}}U_{4,g}(z) &= \frac{2}{N}\bigg(\frac{\sqrt{\lambda }}{ (z-1)^2}-\frac{3 z}{ (z-1)^4}+ \frac{z}{(z-1)^{4}}\Big[\frac{19 \lambda ^{3/2}}{16 N^2}
                                     +\frac{9 \lambda ^3}{64 N^4} -\frac{21 \lambda ^{9/2}}{2560 N^6}-\frac{391 \lambda ^6}{28672 N^8}\Big]\bigg) + \dots ~,\nonumber \\
  \sum_{g}\frac{1}{N^{2g-1}}U_{5,g}(z) &= \frac{5}{2N}\bigg(\frac{\sqrt{\lambda }}{(z-1)^2}-\frac{3 z}{ (z-1)^4} + \frac{z}{(z-1)^{4}}\Big[\frac{25 \lambda ^{3/2}}{16 N^2}
                                      +\frac{5 \lambda ^3}{16 N^4}+\frac{33 \lambda ^{9/2}}{640 N^6}\Big]\bigg) + \dots~.
\ea
As a result of this, the dependence on $J$ in $\frac{\vev{\mathcal{W}^{n}:\tr M^{j}:}}{\vev{\mathcal{W}^{n}}}$ is much simpler for $n>1$
than in the $n=1$ case, \cf (\ref{1.7}). Indeed, from the above, it has to be proportional to 
\ba
  \la{7.24}
 \mathop{\Res}_{z=0}\frac{dz}{(z-1)^{4} z^{J-1}} &= \frac{1}{6} J \left(J^2-1\right) ~.
\ea
This means in that the structure of large tension limit of $\frac{\vev{\mathcal{W}^{n} \mathcal{O}_{J}} }{\vev{\mathcal{W}^{n}}}$ is given by \eqref{1.11}.
The first few terms of $\mathsf{H}_{n}(x)$  can be calculated from \eqref{7.23} and read
\ba
  \la{7.25}
  \mathsf{H}_{2}(x) &= \frac{7 x}{24 \pi }-\frac{x^2}{12 \pi }+\frac{x^3}{20 \pi }-\frac{37 x^4}{1260 \pi }+\frac{13 x^5}{756 \pi }-\frac{299 x^6}{29700 \pi } +\dots~,\nonumber \\
  \mathsf{H}_{3}(x) &= \frac{13 x}{24 \pi }+\frac{x^2}{6 \pi }-\frac{17 x^3}{30 \pi }+\frac{41 x^4}{70 \pi }-\frac{x^5}{189 \pi} + \dots ~,\nonumber \\
  \mathsf{H}_{4}(x) &= \frac{19 x}{24 \pi }+\frac{3 x^2}{4 \pi }-\frac{7 x^3}{20 \pi }-\frac{391 x^4}{84 \pi } + \dots  ~,\nonumber \\
  \mathsf{H}_{5}(x) &= \frac{25 x}{24 \pi }+\frac{5 x^2}{3 \pi }+\frac{11 x^3}{5 \pi } + \dots  ~.
\ea

\subsection{Relating \texorpdfstring{$\mathsf H_{n}$}{Hn} to \texorpdfstring{$\mathsf W_{n}$}{Wn}}
\la{sec:H-W}

The discussion in previous section has led to the expansion (\ref{7.25}) for the scaling functions $\mathsf H_{n}$.
Most importantly, we could prove the general structure (\ref{1.11}), with its peculiar dependence on the $J$ parameter.
In this section we show how this can be exploited to express $\mathsf H_{n}$ in terms of $\mathsf W_{n}$. To this aim we 
take $J=2$ in the topological recursion result (\ref{1.11}) and write
\be
\frac{\vev{\mc W^{n}\mc O_{2}}}{\vev{\mc W^{n}}}  \LT \pi\,n\,(T+3\,\mathsf{H}_{n}).
\ee
The l.h.s. may be traded for a logarithmic derivative of $\vev{\mc W^{n}}$ due to the matrix model identity 
\be
\la{7.27}
\vev{\mc W^{n}\, \mc O_{2}} = \lambda\frac{d}{d\lambda}\vev{\mc W^{n}}.
\ee
Hence we have 
\be
\lambda\frac{d}{d\lambda}\log \vev{\mc W^{n}} \LT \pi\,n\,(T+3\,\mathsf{H}_{n}),
\ee
and a short calculation gives the  relation 
\be
\la{7.29}
\mathsf{H}_{n}(x) = \frac{x}{2\pi}\,\bigg[\frac{1}{12}+\frac{1}{n}\,\frac{d}{dx}\log \mathsf{W}_{n}(x)\bigg].
\ee
Replacing $\mathsf{W}_{n}$ by its evaluation by means of  (\ref{6.20}) and using the series expansion (\ref{5.14}),
we get  
\ba
\la{7.30}
\mathsf{H}_{n}(x) &=
\frac{-5+6 n  }{24 \pi }\,x+\frac{(-1+n) (-5+2 n)}{12 \pi }\,x^{2}+
\frac{(-1+n) (133-95 n+15 n^2) }{60 \pi }\,x^{3}\lp
+\frac{(-1+n) 
(-24159+23611 n-7035 n^2+630 n^3) }{1260 \pi }\,x^{4}+\cdots\ , 
\ea
in agreement with (\ref{7.25}). Of course, the exact determination of $\mathsf W_{n}$ by Toda recursion means that we can 
provide easily all order expansion of the $\mathsf H_{n}$ function by means of (\ref{7.29}).

\subsection{A few sample calculations} 

Let us give some  examples of (\ref{1.12}) by explicit computations. For $n=2$ we need 
the explicit exact  expansion 
\ba
&\frac{1}{N^{2}} \vev{\mc W^{2}} = \Big[\frac{2}{\sqrt\lambda}\,I_{1}\Big]^{2}
+\frac{\sqrt\lambda}{2N^{2}}\Big[
I_{0}\,I_{1}+\frac{1}{6}\,I_{1}\,I_{2}
\Big]
+\frac{1}{N^{4}}\Big[
\frac{37\lambda^{2}}{2304}\,I_{0}^{2}-\frac{\sqrt\lambda(24+131\lambda)}{2880}\,I_{0}\,I_{1}  \lp  
+\frac{192+332\lambda+185\lambda^{2}}{11520}\,I_{1}^{2}
\Big]  +\frac{1}{N^{6}}\Big[
-\frac{\lambda ^2 (62+37 \lambda )}{23040}\,I_{0}^{2}  \lp 
+\frac{\sqrt{\lambda } (23040+56160 \lambda +40920 
\lambda ^2+6209 \lambda ^3)}{5806080}\,I_{0}\,I_{1} \lp
-\frac{92160+111168 \lambda +85440 \lambda ^2+24857 
\lambda ^3}{11612160}\,I_{1}^{2}
\Big]+\mc O\Big(\frac{1}{N^{8}}\Big).
\ea
Using (\ref{7.27}) we work out the case $(n,J) = (2,2)$
\be
\la{7.32}
\frac{\vev{\mc W^{2}\, \mc O_{2}}}{\vev{\mc W^{2}}} =  \sqrt\lambda+\cdots+\frac{1}{N^{2}}\bigg[\frac{7}{32}\lambda^{3/2}+\cdots\bigg]+\frac{1}{N^{4}}\bigg[-\frac{1}{128}\lambda^{3}
+\cdots\bigg]+\frac{1}{N^{6}}\bigg[\frac{3}{5120}\lambda^{9/2}+\cdots\bigg]+\cdots.
\ee
Comparing with (\ref{1.11}) gives the first terms
\be
\la{7.33}
\mathsf{H}_{2}(x) = \frac{1}{6\pi}\bigg(\frac{7}{4}x-\frac{1}{2}x^{2}+\frac{3}{10} x^{3}+\cdots\bigg),
\ee
in agreement with (\ref{7.30}). In this case we can give the exact expression in a reasonable compact form using the first equation in (\ref{5.16})
\be
\la{7.34}
\mathsf{H}_{2}(x) = \frac{3}{2}+2\,x-\frac{3}{2+e^{x/2}\sqrt{2\pi x}\,\text{erf}\big(\sqrt\frac{x}{2}\big)}.
\ee
A similar calculation can be repeated for $n=3$. In this case we have 
\ba
& \frac{1}{N^{3}}  \vev{\mc W^{3}} = 
 \Big[\frac{2}{\sqrt\lambda}\,I_{1}\Big]^{3}+\frac{1}{N^{2}}\,\Big(
\frac{13}{4}\,I_{1}^{2}\,I_{0}-\frac{1}{2\sqrt\lambda}\,I_{1}^{3}
\Big) \lp+\frac{1}{N^{4}}\,\Big[
\frac{193}{384}\lambda^{3/2}I_{0}^{2}I_{1}-\frac{6+79\lambda}{240}\,I_{0}I_{1}^{2}+\frac{192+592\lambda+845\lambda^{2}}{3840\sqrt\lambda}I_{1}^{3}
\Big]\notag \\
&  +\frac{1}{N^{6}}\Big[
\frac{2557 \lambda ^3 }{110592}\,I_{0}^{3}-\frac{\lambda 
^{3/2} (1776+7865 \lambda )}{92160}\,I_{0}^{2}\,I_{1}  +\frac{92160+474624 \lambda +878688 
\lambda ^2+572537 \lambda ^3}{7741440}\,I_{0}\,I_{1}^{2} \lp 
-\frac{23040+46944 \lambda +64396 \lambda ^2+52073 
\lambda ^3}{967680 \sqrt{\lambda }}\,I_{1}^{3}
\Big]+\cdots.
\ea
This gives
\be
\frac{\vev{\mc W^{3}\, \mc O_{2}}}{\vev{\mc W^{3}}} =  \frac{3}{2}\sqrt\lambda+\cdots+\frac{1}{N^{2}}\bigg[\frac{39}{64}\lambda^{3/2}+\cdots\bigg]+\frac{1}{N^{4}}\bigg[\frac{3}{128}\lambda^{3}
+\cdots\bigg]+\frac{1}{N^{6}}\bigg[-\frac{51}{5120}\lambda^{9/2}+\cdots\bigg]+\cdots.
\ee
Comparing with (\ref{1.11}) we obtain
\be
\mathsf{H}_{3}(x) = \frac{1}{9\pi}\,\bigg(\frac{39}{8}\,x+\frac{3}{2}\,x^{2}-\frac{51}{10}\,x^{3}+\cdots\bigg),
\ee
in agreement with (\ref{1.12}). As in (\ref{7.34}), one can give a closed formula for this function in terms of the special error and Owen-T functions.
As a final check, probing the peculiar simple $J$ dependence in (\ref{1.11}), we consider the case $(n,J) = (2,3)$. To analyze this case by expansion of exact expressions at finite $\lambda$ we need
the Bessel function expansion of $\vev{\mc W^{2} \mc O_{3}}$ where $\mc O_{3} = \frac{N}{2}\sqrt\frac{\pi}{2}\,:\tr M^{3}:$ and $:\tr M^{3}: = \tr M^{3}-3 \tr M$.
By matching a large number of weak coupling
perturbative coefficients, we find 
{\small
\ba
\la{7.38}
& \vev{\mc W^{2}\,\mc O_{3}} = N^{2}\sqrt\frac{\pi}{2}\,\bigg\{-\frac{24 I_0 I_1}{\lambda }+\frac{6 (8+\lambda ) I_1^2}{\lambda ^{3/2}}\lp
+\frac{1}{N^{2}}\bigg[
-\frac{1}{4} \sqrt{\lambda } I_0^2+\frac{1}{8} (8+7 \lambda ) I_0 
I_1-\frac{I_1^2}{\sqrt{\lambda }}
\bigg] \lp
+\frac{1}{N^{4}}\bigg[
\frac{\sqrt{\lambda } (192+48 \lambda +185 \lambda ^2) 
I_0^2}{7680}+\frac{(-192-72 \lambda +239 \lambda ^2) I_0 
I_1}{1920}+\frac{(768+384 \lambda -240 \lambda ^2+185 \lambda ^3) 
I_1^2}{7680 \sqrt{\lambda }}
\bigg] \lp
+\frac{1}{N^{6}}\bigg[
\frac{\sqrt{\lambda } (-5760-1440 \lambda -528 \lambda ^2+1939 
\lambda ^3) I_0^2}{483840}+\frac{(184320+69120 \lambda +8544 \lambda 
^2-39902 \lambda ^3+6209 \lambda ^4) I_0 
I_1}{3870720}\lp
+\frac{(-368640-184320 \lambda -6144 \lambda ^2+41128 
\lambda ^3+24815 \lambda ^4) I_1^2}{7741440 \sqrt{\lambda }}
\bigg] +\cdots\bigg\}.
\ea
}
This gives
\be
\frac{\vev{\mc W^{2}\,\mc O_{3}}}{\vev{\mc W^{2}}} = \sqrt\frac{\pi}{2}\bigg\{
 \frac{3}{2}\sqrt\lambda+\cdots+\frac{1}{N^{2}}\bigg[\frac{7}{8}\lambda^{3/2}+\cdots\bigg]+\frac{1}{N^{4}}\bigg[-\frac{1}{32}\lambda^{3}
+\cdots\bigg]+\frac{1}{N^{6}}\bigg[\frac{3}{1280}\lambda^{9/2}+\cdots\bigg]+\cdots\bigg\}.
\ee
This expansion should be compared with the $n=2$ $J=3$ case of (\ref{1.12}), \ie 
\ba
3\,\bigg(\frac{\pi}{2}\bigg)^{3/2}\,2\,(T+8\mathsf{H}_{2}(x)) = \sqrt\frac{\pi}{2}\bigg[
 \frac{3}{2}\times 2\pi T+7\,x-2\,x^{2}+\frac{6}{5}\,x^{3}+\cdots\bigg],
\ea
and indeed we find that this is equivalent to the previous expansion (\ref{7.33}). 

\section*{Acknowledgments}

MB and AH are supported by the INFN grant GSS (Gauge Theories, Strings and Supergravity). 

\appendix

\section{Toda recursion for correlators with chiral primaries}
\la{app:toda-chiral}

The genus expansion of the ratio  $\vev{\mc W \,\mc O_{2}}/\vev{\mc W}$ may be computed by (\ref{7.27}) in terms of $\vev{\mc W}$. Alternatively, it is equivalent
to use the integral representation (\ref{1.10}) derived in \cite{Okuyama:2006jc}. Here, we want to show how such correlators may be treated by 
Toda recursion, as an illustration, generalizing the treatment in App. B.3 of \cite{Beccaria:2020ykg}.
From
\be
e_{N}(x,y) = \vev{\tr e^{x\,\sqrt\frac{4N}{\lambda}M}\tr e^{y \,\sqrt\frac{4N}{\lambda}M}}-\vev{\tr e^{x\,\sqrt\frac{4N}{\lambda}M}}\vev{\tr e^{y\,\sqrt\frac{4N}{\lambda}M}},
\ee
we have 
\be
\la{A.2}
\frac{\left. \partial_{x}^{2}e_{N}\left(x,\sqrt\frac{\lambda}{4N}\right)\right|_{x=0}}{e_{N}\left(\sqrt\frac{\lambda}{4N}\right)} = \frac{4N}{\lambda}\,\frac{\vev{\mc W\,:\tr M^{2}:}}{\vev{\mc W}} = 
2\,\frac{\vev{\mc W\,:\tr a^{2}:}}{\vev{\mc W}},
\ee
where 
\be
M = \sqrt\frac{\lambda}{2N}\,a    \qquad\qquad \bigg[\tr e^{M} e^{-\frac{2N}{\lambda}\tr M^{2}} = \tr e^{\sqrt\frac{\lambda}{2N}\,a}\,e^{-\tr a^{2}}\bigg],
\ee 
to make contact with the expressions in \cite{Beccaria:2020ykg}.
The relevant Toda equation is  (\ref{6.3}). Taking two  derivatives involves the auxiliary quantity
\be
\la{A.4}
\frac{\left. \partial_{x}e_{N}\left(x,\sqrt\frac{\lambda}{4N}\right)\right|_{x=0}}{e_{N}\left(\sqrt\frac{\lambda}{4N}\right)}  = \sqrt\frac{4N}{\lambda}\,\frac{\vev{\mc W\,\tr M}}{\vev{\mc W}} = 
\sqrt{2}\,\frac{\vev{\mc W\,\tr a}}{\vev{\mc W}}.
\ee
To continue, we need the correct Ansatz for the r.h.s. of (\ref{A.2}) and (\ref{A.4}) at large tension. This is 
\be
\frac{\vev{\mc W\,:\tr a^{J}:}}{\vev{\mc W}} = N^{\frac{J}{2}-1}\sqrt\lambda \, C_{J}\left(\zeta\right),\qquad \zeta = \frac{\gs^{2}}{T^{2}} = \frac{\lambda}{4\,N^{2}}.
\ee
The Toda recursion takes the form 
\ba
\la{A.6}
& 4\,\sqrt{N(N-1)}\sqrt\zeta\,C_{2}\left(\frac{N}{N-1}\,\zeta\right) \frac{e_{N-1}(\sqrt{N\zeta})}{e_{N}(\sqrt{N\zeta})}
+4\,\sqrt{N(N+1)}\sqrt\zeta\,C_{2}\left(\frac{N}{N+1}\,\zeta\right) \frac{e_{N+1}(\sqrt{N\zeta})}{e_{N}(\sqrt{N\zeta})} \lp
-4N\sqrt\zeta(2+\zeta)C_{2}(\zeta)-8\sqrt 2\zeta C_{1}(\zeta)+2\zeta=0
\ea
The expansion at large $N$ with fixed $\zeta$ require to study the asymptotic behaviour of $e_{N}(\sqrt{N} \mu)$ at fixed $\mu$.
Recall that 
\be
\la{A.7}
e_{N}(x) = e^{\frac{x^{2}}{2}}L_{N-1}^{1}(-x^{2}), \qquad e_{N}''(x)+\frac{3}{x}e_{N}'(x)-(4N+x^{2})\,e_{N}(x)=0.
\ee
Setting $x=\sqrt{N} \mu$ and expanding the differential equation gives 
%\footnote{The $N^{-1/2}$ prefactor is fundamental and takes into account 
%the boundary condition $e_{N}(0)=N$. It may be derived by Debye expansion of the Laguerre polynomial.}
\be
e_{N}(\sqrt{N}\mu) = N^{-1/2}\exp\bigg[N\,f_{0}(\mu)+f_{1}(\mu)+\frac{1}{N} f_{2}(\mu)+\cdots\bigg],
\ee
with
\ba
f_{0}(\mu) &= \frac{\mu}{2}\sqrt{\mu^{2}+4}+2 \text{arcsinh}\frac{\mu}{2}, \qquad
f_{1}(\mu) = -\frac{3}{2}\log\mu-\frac{1}{4}\log(\mu^{2}+4).
\ea
This is enough to derive the relevant terms in the expansion 
\be
\frac{e_{N\pm 1}(\sqrt{N\zeta})}{e_{N}(\sqrt{N\zeta})} = e^{\pm 2 \text{arcsinh}\frac{\sqrt\zeta}{2}}\bigg[1+\frac{1}{N}\frac{\pm (2+\zeta)-\sqrt\zeta\sqrt{4+\zeta}}{2(4+\zeta)}+\mc O\left(\frac{1}{N^{2}}\right)\cdots\bigg].
\ee
Using this in the expansion of (\ref{A.6}) gives
\be
C_{2}'(\zeta)-\frac{1}{2\zeta}C_{2}(\zeta)+\frac{-1+4\sqrt 2\,C_{1}(\zeta)}{2\zeta\sqrt{4+\zeta}}=0.
\ee
It is easy to check that $C_{1}(\zeta) \equiv \frac{1}{2\sqrt 2}$, so that 
\be
C_{2}(\zeta) = \frac{1}{4}\sqrt{4+\zeta}+k\,\sqrt\zeta,
\ee
where $k$ is a constant that we set to zero by analyticity. The result agrees with \cite{Beccaria:2020ykg}, see Eqs.(2.34, 2.35, 2.40) there.

\section{The polynomial $P_{k}$ for $k=11, \dots, 20$}
\la{app:ppp}

The polynomials $P_{k}(n)$ have been defined in (\ref{5.18}) and their expression for $k$ up to 10 have been 
given in (\ref{5.19}) and (\ref{uu1}). The expressions for $k=11, \dots, 20$  are given below.
\begin{align}
  P_{11}&=\frac{4199 n^{10}}{33}-\frac{87577 n^9}{11}+\frac{21375883 n^8}{99}-\frac{209156735279 n^7}{62370}+\frac{5161998742529 n^6}{155925}\nonumber \\
        &-\frac{101671734522896 n^5}{467775}+\frac{77972872201319 n^4}{81081}-\frac{120508596162974836 n^3}{42567525}\nonumber \\
        &+\frac{2558119973701481627 n^2}{482431950}-\frac{335724211283681053843 n}{58925616750}+\frac{156764630068025273339}{58925616750} ~, \nonumber \\
  P_{12}&=\frac{2261 n^{11}}{6}-\frac{965975 n^{10}}{36}+\frac{10060585 n^9}{12}-\frac{172514522407 n^8}{11340}+\frac{30353622743141 n^7}{170100}\nonumber \\
        &-\frac{483860927926589 n^6}{340200}+\frac{40171770380189773093 n^5}{5108103000}-\frac{51369864627494293667 n^4}{1702701000}\nonumber \\
        &+\frac{102472685620882550488819 n^3}{1302566265000}-\frac{574763444256918348659002 n^2}{4331032831125}\nonumber \\
        &+\frac{393843049171726870141793 n}{3024848326500}-\frac{98527083191201048311039181}{1753202090039400}  ~, \nonumber \\
  P_{13}&=\frac{14858 n^{12}}{13}-\frac{3561572 n^{11}}{39}+\frac{1890539597 n^{10}}{585}-\frac{824252140702 n^9}{12285}
          +\frac{11192496489158 n^8}{12285}\nonumber \\
        &-\frac{942785947894636 n^7}{110565}+\frac{14402080240084077257 n^6}{255405150}-\frac{203271501770547142769 n^5}{766215450}\nonumber \\
        &+\frac{9555769999611687383321 n^4}{10854718875}-\frac{11588871456123457506361969 n^3}{5774710441500}\nonumber \\
        &+\frac{2838595185137082339418889941 n^2}{952827222847500}-\frac{51452536482602877650860472 n}{19922751023175}\nonumber \\
        &+\frac{3608383690059034257720530341}{3652504354248750} ~, \nonumber \\
  P_{14}&=\frac{74290 n^{13}}{21}-\frac{2193609 n^{12}}{7}+\frac{3898791707 n^{11}}{315}-\frac{54577362356 n^{10}}{189}
          +\frac{62912522293682 n^9}{14175}\nonumber \\
        &-\frac{671658180402872 n^8}{14175}+\frac{212496790875244595807 n^7}{589396500}-\frac{30206956766158029102587 n^6}{15324309000}\nonumber \\
        &+\frac{2004508256000686158596273 n^5}{260513253000}-\frac{726580590679310973992290361 n^4}{34648262649000}\nonumber \\
        &+\frac{2908859723836817593513494211 n^3}{76226177827800}-\frac{46704914627229276739182458986 n^2}{1095751306274625}\nonumber \\
        &+\frac{6980915896986492598862795960273 n}{284895339631402500}-\frac{745725550174967494764854054797}{170937203778841500} ~, \nonumber \\
  P_{15}&=\frac{22287 n^{14}}{2}-\frac{97455637 n^{13}}{90}+\frac{471242699 n^{12}}{10}-\frac{7296074705 n^{11}}{6}
          +\frac{1762900954888741 n^{10}}{85050}\nonumber \\
        &-\frac{229802660901854531 n^9}{935550}+\frac{29028607670070832589 n^8}{14033250}-\frac{4742825903093184375173 n^7}{383107725}\nonumber \\
        &+\frac{116822823990093464083591 n^6}{2277213750}-\frac{17658116093652216575266660279 n^5}{129930984933750}\nonumber \\
        &+\frac{3019686155476492526344171776962 n^4}{17865510428390625}+\frac{59133660041425626175154220297689 n^3}{328725391882387500}\nonumber \\
        &-\frac{40902420138582238361745904216740547 n^2}{38460870850239337500}+\frac{5184745996967930004978479698356554 n}{3205072570853278125}\nonumber \\
        &-\frac{8779046910792332017382938446939173}{9783905742604743750} ~, \nonumber \\
  P_{16}&=\frac{570285 n^{15}}{16}-\frac{361327963 n^{14}}{96}+\frac{256924672387 n^{13}}{1440}-\frac{151766993879233 n^{12}}{30240}
          +\frac{938030587963673 n^{11}}{10080}\nonumber \\
        &-\frac{5938730301115621841 n^{10}}{4989600}+\frac{44402659987134144011 n^9}{4191264}-\frac{776491656618909257262217 n^8}{12259447200}\nonumber \\
        &+\frac{675670971369618816638500351 n^7}{3126159036000}+\frac{65379058846039414975807037 n^6}{24457597164000}\nonumber \\
        &-\frac{101233943008886150795055382519397 n^5}{22867853348340000}+\frac{13481261415837935476957015174331149 n^4}{525960627011820000}\nonumber \\
        &-\frac{11425876041556661448955324338319179059 n^3}{143587251174226860000}
          +\frac{24484874250586308873555049657424059483 n^2}{165677597508723300000}\nonumber \\
        &-\frac{9643906311965688880889035950642704853709 n}{62460454260788684100000}
          +\frac{1802190759535306926724608738299225695879}{25816987761125989428000} ~, \nonumber \\
  P_{17} &= \frac{1964315 n^{16}}{17}-\frac{39445288 n^{15}}{3}+\frac{60467476843 n^{14}}{90}-\frac{115225766352631 n^{13}}{5670}
           +\frac{253254621872909 n^{12}}{630}\nonumber \\
        &-\frac{5026692736385852671 n^{11}}{935550}+\frac{13913460599257648861603 n^{10}}{294698250}-\frac{122986340264675433569593 n^9}{547296750}\nonumber \\
        &-\frac{2027150038463820624702058 n^8}{6512831325}+\frac{1295892244324167362329025833687 n^7}{86620656622500}\nonumber \\
        &-\frac{57860508893407313308984093326059 n^6}{433103283112500}+\frac{1873177157824570617487083559682467 n^5}{2629803135059100}\nonumber \\
        &-\frac{27442715333021284930815598305084080569 n^4}{10769043838067014500}
          +\frac{1970688218629393121039936070672626597 n^3}{318610764439852500}\nonumber \\
        &-\frac{452466495402865130794144735387180208039 n^2}{46198560843778612500}
          +\frac{883182453823558336713057441231757494529669 n}{96813704104222460355000}\nonumber \\
        &-\frac{4841495934337212459267479837954600070947}{1280604551643154237500} ~, \nonumber \\
 P_{18} &=\frac{1137235 n^{17}}{3}-\frac{2490913949 n^{16}}{54}+\frac{679389195281 n^{15}}{270}-\frac{228968967202858 n^{14}}{2835}\nonumber \\
        &+\frac{2027925651493864 n^{13}}{1215}-\frac{33176267023544737 n^{12}}{1485}+\frac{18475026421421474122847 n^{11}}{109459350}\nonumber \\
        &+\frac{32427853962749456647193 n^{10}}{1532430900}-\frac{141562392656331388912303109 n^9}{7662154500}\nonumber \\
        &+\frac{4407051037515992567242285006784 n^8}{16705412348625}-\frac{12998232848938898943788373084481291 n^7}{5846894322018750}\nonumber \\
        &+\frac{1920684931913550119678771491316267836 n^6}{147926426347074375}
          -\frac{5256663538285399899069978418495823491471 n^5}{96152177125598343750}\nonumber \\
        &+\frac{16669479504189008529752425764572885471 n^4}{100332706565841750}\nonumber \\
        &-\frac{16671094370233566929477876372974768312804309 n^3}{46845340695591513075000}\nonumber \\
        &+\frac{100433411639637043975757841635773385645900627 n^2}{197213100953045752575000}\nonumber \\
        &-\frac{43847679911162088556712200641949149804409228141 n}{100578681486053333813250000}\nonumber \\
        &+\frac{1986475693390240135403726810007174861110233163}{11807062609232347882425000} ~, \nonumber \\
 P_{19} &= \frac{23881935 n^{18}}{19}-\frac{9257432861 n^{17}}{57}+\frac{8012771594381 n^{16}}{855}-\frac{754041922791055 n^{15}}{2394}\nonumber \\
        &+\frac{595825494782601934 n^{14}}{89775}-\frac{35206774729032743512 n^{13}}{423225}+\frac{188028571662687223154182 n^{12}}{577702125}\nonumber \\
        &+\frac{356196397870788389963150374 n^{11}}{36395233875}-\frac{26744467018972588685277494738 n^{10}}{109185701625}\nonumber \\
        &+\frac{123556902465696451688500797402787 n^9}{38979295480125}-\frac{164109225413861098250930345185156307 n^8}{5846894322018750}\nonumber \\
        &+\frac{76973564904301165571543256581512685027 n^7}{422646932420212500}\nonumber \\
        &-\frac{1190462126419784655208489507867266436505147 n^6}{1346130479758376812500}\nonumber \\
        &+\frac{4326837524673482906294836489485405025912703 n^5}{1346130479758376812500}\nonumber \\
        &-\frac{31045843858795962745573700604629252477544851 n^4}{3603487745814731775000}\nonumber \\
        &+\frac{175717184221761997495399027443863716431266200813 n^3}{10649507451464470639050000}\nonumber \\
        &-\frac{11573462752723826199667926549477988549478191522987 n^2}{543124880024688002591550000}\nonumber \\
        &+\frac{2046376907527838309214791130013642260390433408739 n}{123437472732883636952625000}\nonumber \\
        &-\frac{77748831089365050929680848793957593192606798599}{13290754339228476253893750} ~, \nonumber \\
 P_{20} &= \frac{8415539 n^{19}}{2}-\frac{34412953709 n^{18}}{60}+\frac{6246399383567 n^{17}}{180}-\frac{13656186763905823 n^{16}}{11340}\nonumber \\
        &+\frac{1422187058979010861 n^{15}}{56700}-\frac{45793882693341165107 n^{14}}{178200}-\frac{1234280007795319427540237 n^{13}}{729729000}\nonumber \\
        &+\frac{357438485323679466115258847 n^{12}}{3064861800}-\frac{545681680470218713526170387867 n^{11}}{229864635000}\nonumber \\
        &+\frac{138672908598687244744136052146053 n^{10}}{4476091347000}-\frac{376350262572215475642638310861327499277 n^9}{1286316750844125000}\nonumber \\
        &+\frac{120017260584537152890452186439443309163 n^8}{57671121382875000}\nonumber \\
        &-\frac{77980193487962152970270963662541582475107 n^7}{6903233229530137500}\nonumber \\
        &+\frac{9322371447260705245309378650978156112491889 n^6}{199426737741981750000}\nonumber \\
        &-\frac{22752890372440420457183359365326085913312317041 n^5}{156151135651971710250000}\nonumber \\
        &+\frac{5928819160164250201043568590148571162631031954661 n^4}{17749179085774117731750000}\nonumber \\
        &-\frac{10281819624013490020658374119628597074265484293037677 n^3}{19009370800864080090704250000}\nonumber \\
        &+\frac{10982787902444574382251854966765664605681046163557857 n^2}{19009370800864080090704250000}\nonumber \\
        &-\frac{1795787195145485845354622910580596712088982772028004517 n}{5011345377377793113911907906250}\nonumber \\
        &+\frac{9747831016605627462652486155608940497922541559209369}{102796828253903448490500675000}
\end{align}

\section{Some details about topological recursion at large tension}
\la{app:toprec-sub}

Here we summarize some details about topological recursion that are relevant to the strong coupling limit of correlation functions studied in the main text. Our
presentation will be for the Gaussian matrix model although most of the statements have straightforward generalizations to a general genus $0$ spectral curve. See
\cite{Eynard:2008we,Eynard:2015aea} for pedagogical details and general treatment.

\subsection{Spectral curve, resolvents, and residues}
\la{app:sec:spectral-curve}

For the Gaussian matrix model the spectral curve is a two-sheeted cover of the complex plane. \footnote{This holds more generally in 1-cut cases, not necessarily Gaussian. 
The spectral curve has genus $s$ when the large $N$ limit is associated with $s+1$ disconnected cuts.} 
The two sheets are glued along the cut on which the eigenvalues condense in the large $N$ limit.
The coordinate $z$ defined in \eqref{2.3} maps these two sheets to the Riemann sphere as shown in Fig.~\ref{fig:1}.
A generic value of $x$ has two preimages since $x(z) = x\left(\frac{1}{z}\right)$, if $\abs{z} \ne 1$ then for one of these preimages $\abs{z} > 1$ and for other
$\abs{z} < 1$. These are the two sheets which have been mapped to the exterior and interior respectively of unit circle on the $z$-plane. Let's now focus on
the unit circle itself on which we write $z = \exp(i t)$. Then $x(z(t)) = 2\cos t$. So as $z$ goes from $0$ to $\pi$, $x(z(t))$ goes from $2$ to $-2$. This is one
copy of the cut while the other copy is corresponds to $t$ going from $\pi \to 2\pi \sim  0$. The two copies of the cut are joined at $z=1$ and $z=-1$ which correspond
to $x=2$ and $x=-2$ i.e the end points of the cut. These are the only two values of $x$ which have a single preimage. These are the zeroes of the differential
$d x$. Lastly, notice that although $y$ is not a single valued function of $x$, it is a single valued function of $z$. Note that the unit circle is also the
contour for the saddle point approximation, the saddle point integral is actually done over a double copy of the cut.
\begin{figure}[htbp]
    \centering
    \begin{subfigure}[t]{0.45\textwidth}
        \centering
%        \begin{tikzpicture}
%            \draw[draw=none] (-3,-2) rectangle (3,2);
%            \draw  (-2,0) node [draw=black, cross , inner sep = 2pt , fill opacity=1 ](a){};
%            \draw  (2,0) node [draw=black, cross , inner sep = 2pt , fill opacity=1 ](b){};
%            \draw  (-2,0) node [draw=black, circle , inner sep = 2pt , fill opacity=0 ](a){};
%            \draw  (2,0) node [draw=black, circle , inner sep = 2pt , fill opacity=0 ](b){};
%            \draw[color = darkred] (a) -- (b);
%            \draw [darkcyan,ultra thick] plot [smooth, tension=1] coordinates {(0,0)(0.5,0.75) (1,1)};
%            \draw [darkcyan,densely dashed,  ultra thick] plot [smooth, tension=1] coordinates {(0,0)(-0.5,-0.75) (-1,-1)};
%            \node[right] at (1,1) {B};
%            \node[left] at (-1,-1) {A};
%        \end{tikzpicture}
        \begin{tikzpicture}
            \draw[draw=none] (-3,-2) rectangle (3,2);
            \draw (-3,1)--(-2,2)--(3,2)--(2,1)--(-3,1);
            \fill[color=yellow,opacity=0.1] (-3,1-3)--(-2,2-3)--(3,2-3)--(2,1-3)--(-3,1-3);
            \draw (-3,1-3)--(-2,2-3)--(3,2-3)--(2,1-3)--(-3,1-3);
            \draw[color=darkred] (-1,1.5)--(1,1.5);
            \draw[color=darkred] (-1,1.5-3)--(1,1.5-3);
            \draw  (-1,1.5) node [draw=black, cross , inner sep = 2pt , fill opacity=1 ](a){};
            \draw  (1,1.5) node [draw=black, cross , inner sep = 2pt , fill opacity=1 ](b){};
            \draw  (-1,1.5) node [draw=black, circle , inner sep = 2pt , fill opacity=0 ](a){};
            \draw  (1,1.5) node [draw=black, circle , inner sep = 2pt , fill opacity=0 ](b){};
            \draw  (-1,1.5-3) node [draw=black, cross , inner sep = 2pt , fill opacity=1 ](a){};
            \draw  (1,1.5-3) node [draw=black, cross , inner sep = 2pt , fill opacity=1 ](b){};
            \draw  (-1,1.5-3) node [draw=black, circle , inner sep = 2pt , fill opacity=0 ](a){};
            \draw  (1,1.5-3) node [draw=black, circle , inner sep = 2pt , fill opacity=0 ](b){};
            \draw[darkcyan, thick] plot [smooth, tension=1] coordinates {(0,1.5)(0.3,1.8)(0.5,1.9)};
            \draw[darkcyan, thick] plot [smooth, tension=1] coordinates {(0,1.5-3)(-0.3,-1.8)(-0.5,-1.9)};
            \draw[darkcyan, thick, densely dashed] (0,1.5)--(0,1.5-3);
            \node[right] at (0.45,1.83) {$B$};
            \node[left] at (-0.5,-1.8) {$A$};
            \fill[color=cyan, opacity=0.05] (-1,1.5)--(1,1.5)--(1,1)--(-1,1);
            \fill[color=cyan, opacity=0.1] (-1,1)--(1,1)--(1,1.5-3)--(-1,1.5-3);
        \end{tikzpicture}
        \subcaption*{$x$-plane (two sheets)}
    \end{subfigure}
    \begin{subfigure}[t]{0.45\textwidth}
        \centering
        \begin{tikzpicture}
            \draw[draw=none] (-3,-2) rectangle (3,-2);
	   \fill [color=yellow,opacity=0.1] (0,0) circle (1);
            \draw [color=darkred] (0,0) circle (1);
            \draw  (-1,0) node [draw=black , circle , inner sep = 2pt , fill=white , opacity=1](a){};
            \draw  (-1,0) node [draw=black , cross , inner sep = 2pt , fill=white , opacity=1](a){};
            \draw  (1,0) node [draw=black , circle , inner sep = 2pt , fill=white , opacity=1](b){};
            \draw  (1,0) node [draw=black , cross , inner sep = 2pt  , fill=white , opacity=1](b){};
            \begin{scope}[xshift=0.4cm, yshift=-0.15cm]
	            \draw [darkcyan,  thick] plot [smooth, tension=1] coordinates {(0,1)(0.3,1.38) (0.75,1.5)};
	         	   \draw [darkcyan,  thick] plot [smooth, tension=1] coordinates {(0,1)(-0.3,0.5)(-0.5,0.5)};
	            \node[right] at (0.75,1.5) {$B$};
	            \node[below] at (-0.5,0.45) {$A$};
	   \end{scope}
            \end{tikzpicture}
           \subcaption*{$z$-plane}
    \end{subfigure}
    \caption{    \la{fig:1}
	Illustration of the spectral curve. 
	{\bf Left:} the two $x$-sheets connected by the red cut. The points $A$ and $B$ are in different sheets. 
	{\bf Right:} In the $z$-plane the circle is formed from two copies of the cut and separates the two sheets whose images are the outer/inner parts. In particular, the (image of the ) point $A$ is inside the circle.}
\end{figure}
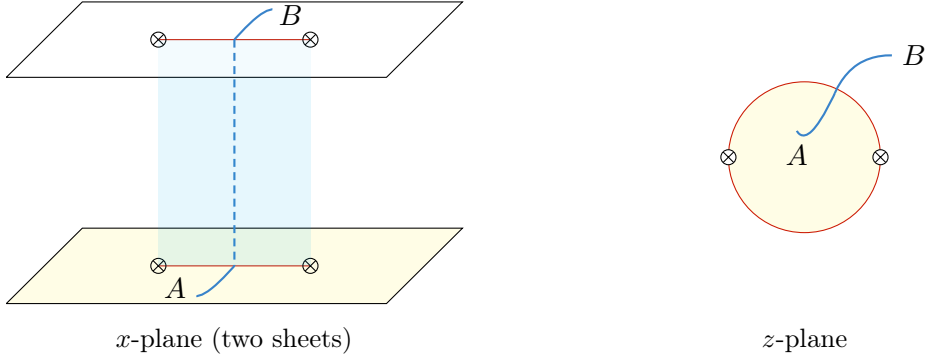

%\subsection{Resolvents and residues}
%\la{app:sec:resolvents-residues}

The resolvents $\omega_{n,g}(z_{1} , \dots , z_{n})$ are all meromorphic multi-differentials on the $z$-plane which poles only at the branch point $z= \pm 1$. One of
results of the topological recursion is the antisymmetry property:
\ba
  \la{B.1}
  \omega_{n,g}\left(\frac{1}{z_{1}} ,z_{2} , \dots , z_{n} \right) &= - \omega_{n,g}(z_{1} , \dots , z_{n}) ~.
\ea
As a consequence,  correlation functions of the polynomials formed from the trace of the matrix $M$ don't receive any
contribution from the poles of the resolvents. This follows from the fact that these matrix observables map to polynomials $f(x(z_{1}), \dots , x(z_{n}) )$. And since
$x(z) = x\left(\frac{1}{z}\right)$, the same is true of $f$ too. Since $z \to \frac{1}{z}$ leaves $\pm 1$ fixed, changing variables to $\frac{1}{z}$, we obtain
\ba
  \la{B.2}
  \mathop{\Res}_{z_{1}=\pm 1}\omega_{n,g}(z_{1} , \dots , z_{n}) f(x(z_{1}), \dots , x(z_{n}) )&= \mathop{\Res}_{z_{1}=\pm 1} \omega_{n,g}\left(\frac{1}{z_{1}} ,z_{2} , \dots , z_{n} \right)f\left(x\left(\frac{1}{z_{1}}\right), \dots , x(z_{n}) \right)\nonumber \\
                                 &= - \mathop{\Res}_{z_{1}=\pm 1}\omega_{n,g}(z_{1} , \dots , z_{n}) f(x(z_{1}), \dots , x(z_{n}) ) = 0 ~.
\ea
As a result the sole contribution to the correlation function of $f$ comes from its own poles, which for a polynomial of $x_{i}$ are at $z(x_{i})=0$ (inside the
contour) and $z(x_{i}) \to \infty$ (outside the contour). These correspond to $x \to  \infty$ in the two sheets. Hence, we can write
\ba
  \la{B.3}
  \oint_{z_{i}} \omega_{n,g}(z_{1} , \dots , z_{n}) f(x(z_{1}), \dots , x(z_{n}) ) &= \mathop{\Res}_{z_{i}=0} \omega_{n,g}(z_{1} , \dots , z_{n}) f(x(z_{1}), \dots , x(z_{n}) ).
\ea
The same logic works for any holomorphic function of $x$ among them the Wilson loop. As we have seen in practice, the contour integral is more
convenient for the strong coupling expansion of Wilson loops while the residue at $0$ is simpler for chiral operators. Nevertheless this vanishing of residues
at $\pm 1$ ensures that there is no ambiguity in the saddle point prescription, since we can smoothly deform the contour past the branch points, as illustrated in Fig.~\ref{fig:2}.
\sidecaptionvpos{figure}{c}
\begin{SCfigure}%[htbp]
    \centering
    % \captionsetup[subfigure]{laformat=empty}
    \begin{subfigure}[t]{0.45\textwidth}
        \centering
        \begin{tikzpicture}
            \draw[draw=none] (-3,-2) rectangle (3,2);
            \draw [color=darkred ] (0,0) circle (1);
            \draw [color=darkcyan , thick] (0,0) circle (1.6);
            \draw  (-1,0) node [draw=black , circle , inner sep = 2pt , fill=white , opacity=1](a){};
            \draw  (-1,0) node [draw=black , cross , inner sep = 2pt , fill=white , opacity=1](a){};
            \draw  (1,0) node [draw=black , circle , inner sep = 2pt , fill=white , opacity=1](b){};
            \draw  (1,0) node [draw=black , cross , inner sep = 2pt , fill=white , opacity=1](b){};
            \draw  (0,0) node [draw=black , circle , inner sep = 2pt , fill=white , opacity=1](c){};
            \draw  (0,0) node [draw=black , cross , inner sep = 2pt , fill=white , opacity=1](c){};
            \draw [color=darkcyan, densely dashed, thick] (a) circle (0.33);
            \draw [color=darkcyan, densely dashed, thick] (b) circle (0.33);
            \draw [color=darkcyan , thick] (c) circle (0.33);
            \fill [color=darkcyan,thick] (0.2,1.7)--(0,1.6)--(0.2,1.5);
            \fill [color=darkcyan,thick] (0.12,0.38)--(0,0.33)--(0.1,0.25)--(0.12,0.38);
        \end{tikzpicture}
        \caption*{$z$-plane}
    \end{subfigure}
%    \begin{subfigure}[t]{0.45\textwidth}
%        \centering
%        \begin{tikzpicture}
%            \draw [draw=none] (-3,-2) rectangle (3,2);
%            \draw  (-2,0) node [draw=red , circle , inner sep = 2.5pt , fill=white , opacity=1](a){};
%            \draw  (-2,0) node [draw=red , cross , inner sep = 2.5pt , fill=white , opacity=1](a){};
%            \draw  (2,0) node [draw=red , circle , inner sep = 2.5pt , fill=white , opacity=1](b){};
%            \draw  (2,0) node [draw=red , cross , inner sep = 2.5pt , fill=white , opacity=1](b){};
%            \draw [color = black , thick , dashed] (a) -- (b);
%            \draw [color=green , thick] (a) circle (0.33);
%            \draw [color=green , thick] (b) circle (0.33);
%            \draw [color=black , thick] (0,0) ellipse (2.5 and 1.25);
%        \end{tikzpicture}
%        \caption{An $x$-sheet}
%    \end{subfigure}
    \caption{
    	\la{fig:2}
    	The contour integral for computing the matrix correlator $f(x)$ in $z$-plane (the big circle) gets contribution only from the pole of
          $f(x(z))$ at $z=0$. The residue at poles of resolvent (dashed circles) vanish. }
          % (b) The same picture in the $x$-sheet, the only contribution to the
          % contour (the ellipse) integral comes from $x \to  \infty$ in the other sheet which is reached via the branch cut (dashed line).}
\end{SCfigure}

\subsection{Topological recursion at subleading order}\la{app:sec:topol-recurs-at}
The coordinate $u$ defined in \eqref{3.13} which is convenient for extending the saddle point approximation to subleading orders can be seen as a
reparameterization of spectral curve as
\ba
  \la{B.4}
  z(u) = \exp\left(2i\arcsin\left(\frac{u}{2}\right)\right) ~.
\ea
This change of variables maps the $z$-plane to a cylinder $u = \phi + i r$ where $\phi$ parameterizes a circle of radius $4$ extending from $-2$ to $2$ while $r$ a
real line. In this manner $u$ is the local complex coordinate on an infinite cylinder. This cylinder is compactified to a sphere by identifying the circle at
$u = i\infty$ with one point and the circle at $u=-i\infty$ with another point. In the $u$-coordinate the branch point are mapped to $0$ and $-2 \sim 2$. In the strong
coupling limit the dominant contribution to the expectation value of Wilson loops comes from $u=0$ while the contour for saddle point integral is the circle $r = 0$.

These coordinates also turns out to be somewhat simpler for carrying out topological recursion. Changing variables, the topological recursion formula \eqref{2.8} becomes
\ba
  \la{B.5}
      \omega_{n,g}(u_{1},\mathbf{u}) &= \mathop{\Res}_{v \to 0, \pm 2}K(u_{1},v) \Big[\omega_{g-1,n+1}\left(v,-v,\mathbf{u}\right) + \sum_{h\le g}\sum_{\mathbf{r}\subset \mathbf{u}}\omega_{h,\abs{\mathbf{r}}}
                                \left(v,\mathbf{r}\right)\omega_{g-h,n-\abs{\mathbf{r}}}\left(-v,\mathbf{u}/\mathbf{r}\right)\Big] ~.
\ea
In terms of these variables the recursion Kernel $K$ is
\ba
  \la{B.6}
  K(u,v) &= -\frac{i\,d u}{2v \sqrt{4-u^2} \left( u^2 - v^2\right)d v} ~.
\ea
Apart from the factor $\sqrt{u^{2}-4}$ which is independent of $v$ and as a result gives an overall multiplicative factor, the kernel is homogeneous in these
coordinates if we only keep the residue at $u=0$. This makes it easier to separate out the contribution of different orders. Indeed defining $\hat{K}(u,v) = -i\,\sqrt{4-u^{2}}K(u,v)$, we see that
\ba
  \la{B.7}
  \mathop{\Res}_{v \to 0}\hat{K}(u,v) \frac{d u}{v^{2k}} &= -\frac{d u}{2 u^{2k+2}} ~, \nonumber \\
 \mathop{\Res}_{v \to 0}\hat{K}(u,v) \frac{d v}{(v-w)^{2}v^{2k}} &= - \frac{d u}{2} \sum_{i=0}^{k}\frac{2i+1}{w^{2i+2}u^{2k-2i+2}} ~.
\ea
So $\hat{K}(u,v)$ uniformly increases the degree of the poles of differential it acts on by $2$. This simplification in the recursion kernel is a trade off due to
the fact that the starting point of the recursion $\omega_{2,0}(u,v)$ is now more complicated being given by
\ba
  \la{B.8}
  w_{2,0}(u,v) &= \frac{8 d u\, d v}{\sqrt{4-u^2} \sqrt{4-v^2} \left(2 u^2+2 v^2-u^2 v^2-u v \sqrt{4-u^2} \sqrt{4-v^2}\right)}\ ,
\ea
and,  for the purposes of carrying out topological recursion, it will be expanded into a double power series easily.
Another simplification is that in these coordinates the antisymmetry property \eqref{B.1} reads
\ba
  \la{B.9}
  \omega_{n,g}(-u_{1} , \dots , u_{n}) &= \omega_{n,g}(u_{1} , \dots , u_{n})\ .
\ea
This means in particular that
\ba
  \la{B.10}
  \omega_{n,g}(u_{1} , \dots , u_{n}) = i^{n} \, d u_{1}\dots  d u_{n}\, f_{n,g}\left(\frac{1}{u_{1}^{2}} , \dots , \frac{1}{u_{n}^{2}}\right),
\ea
for some symmetric polynomials $f_{n,g}$.  As a result the poles encountered in the saddle point integrals are always of even order. Finally, we observe that all $\omega_{g,n}$
computed through the topological recursion have poles of order at least $4$ at $u=0$ and as a result for the first two orders of poles that we need we can
ignore the residues at $\pm 2$ in \eqref{B.5}.

\subsection{Expressions for resolvents at leading and first subleading order of poles}\la{sec:expr-resolv-at}
\la{app:resolvents}

Now we present some of the resolvents needed to compute various explicit expansions presented in the main text (\ref{5.10},~\ref{7.22},~\ref{7.23}). 
We do this by presenting $f_{n,g}\left(\frac{1}{u_{1}^{2}} , \dots , \frac{1}{u_{n}^{2}}\right)$ as defined
above in \eqref{B.10}. These are symmetric polynomials of their arguments $\frac{1}{u_{i}^{2}}$ and to keep the expression relatively compact we present them in terms of elementary symmetric
polynomials, \cf (\ref{4.3}).
 Similarly to our decomposition of $\omega_{n,g} = \hat{\omega}_{n,g} + \delta \omega_{n,g}$ we divide $f_{n,g}$ into leading $\hat{f}_{n,g}$ and subleading
$\delta f_{n,g}$ pieces.

\paragraph{\underline{Leading order $\hat{\omega}_{g,n}$}}

\paragraph{$n=1$}
\ba
  \la{B.11}
  \hat{f}_{1,1} &= \frac{e_1^2}{16}~,\nonumber \\
  \hat{f}_{1,2} &= -\frac{105 e_1^5}{1024} ~,\nonumber \\
  \hat{f}_{1,3} &=\frac{25025 e_1^8}{32768}~,\nonumber \\
  \hat{f}_{1,4} &= -\frac{56581525 e_1^{11}}{4194304}~,\nonumber \\
  \hat{f}_{1,5} &= \frac{58561878375 e_1^{14}}{134217728}~,\nonumber \\
  \hat{f}_{1,6} &= -\frac{193039471750125 e_1^{17}}{8589934592}~,\nonumber \\
  \hat{f}_{1,7} &= \frac{464259929559050625 e_1^{20}}{274877906944}~,\nonumber \\
  \hat{f}_{1,8} &= -\frac{12277353837189093778125 e_1^{23}}{70368744177664} ~.
\ea
\paragraph{$n=2$}
\ba
  \la{B.12}
  \hat{f}_{1,1} &= \frac{5}{32} e_1^2 e_2-\frac{7 e_2^2}{32}~,\nonumber \\
  \hat{f}_{2,2} &= -\frac{1155 e_2 e_1^5}{2048}+\frac{2415 e_2^2 e_1^3}{1024}-\frac{3955 e_2^3 e_1}{2048}~,\nonumber \\
  \hat{f}_{2,3} &= \frac{425425 e_2 e_1^8}{65536}-\frac{3028025 e_2^2 e_1^6}{65536}+\frac{6641635 e_2^3 e_1^4}{65536}-\frac{4582655 e_2^4 e_1^2}{65536}
            +\frac{119665 e_2^5}{16384}~,\nonumber \\
  \hat{f}_{2,4} &= -\frac{1301375075 e_2 e_1^{11}}{8388608}+\frac{1640864225 e_2^2 e_1^9}{1048576}-\frac{23885486625 e_2^3 e_1^7}{4194304}
            +\frac{18836542725 e_2^4 e_1^5}{2097152} \nonumber \\
          &\phantom{abc}-\frac{47950777875 e_2^5 e_1^3}{8388608}+\frac{2103075975 e_2^6 e_1}{2097152}~,\nonumber \\
  \hat{f}_{2,5} &=\frac{1698294472875 e_2 e_1^{14}}{268435456}-\frac{22194951904125 e_2^2 e_1^{12}}{268435456}
            +\frac{56694405142375 e_2^3 e_1^{10}}{134217728}-\frac{142801188354825 e_2^4 e_1^8}{134217728}\nonumber \\
           &\phantom{abc}+\frac{366814122449775 e_2^5 e_1^6}{268435456}-\frac{222563485169025 e_2^6 e_1^4}{268435456}
             +\frac{6268140053175 e_2^7 e_1^2}{33554432}-\frac{113818730025 e_2^8}{16777216}~,\nonumber \\
  \hat{f}_{2,6} &=-\frac{6756381511254375 e_2 e_1^{17}}{17179869184}+\frac{54244091561785125 e_2^2 e_1^{15}}{8589934592}
            -\frac{714860261976485625 e_2^3 e_1^{13}}{17179869184}\nonumber \\
          &\phantom{abc}+\frac{622461720688933125 e_2^4 e_1^{11}}{4294967296}
            -\frac{4917560490983765625 e_2^5 e_1^9}{17179869184}+\frac{2741103726364128125 e_2^6 e_1^7}{8589934592}\nonumber \\
            &\phantom{abc}-\frac{3222371840821508375 e_2^7 e_1^5}{17179869184}+\frac{105576169391317625 e_2^8 e_1^3}{2147483648}
              -\frac{4002195436113875 e_2^9 e_1}{1073741824} ~,\nonumber \\
  \hat{f}_{2,7} &= \frac{19034657111921075625 e_2 e_1^{20}}{549755813888}-\frac{362587004985618538125 e_2^2 e_1^{18}}{549755813888}
            +\frac{2928158800333479838125 e_2^3 e_1^{16}}{549755813888}\nonumber \\
            &\phantom{abc}-\frac{13056036458334748003125 e_2^4 e_1^{14}}{549755813888}
              +\frac{35071726847099021454375 e_2^5 e_1^{12}}{549755813888}\nonumber \\
            &\phantom{abc}-\frac{58109227982455794519375 e_2^6 e_1^{10}}{549755813888}
              +\frac{58386905310562738284375 e_2^7 e_1^8}{549755813888}\nonumber \\
            &\phantom{abc}-\frac{33548300609678088609375 e_2^8 e_1^6}{549755813888}
              +\frac{2434905970723398046875 e_2^9 e_1^4}{137438953472}\nonumber \\
            &\phantom{abc}-\frac{68144533015364430625 e_2^{10} e_1^2}{34359738368}
            +\frac{312467552127355625 e_2^{11}}{8589934592}~.
\ea

\paragraph{$n=3$}
\ba
  \la{B.13}
  \hat{f}_{3,0} &=-\frac{e_3}{2}~,\nonumber \\
  \hat{f}_{3,1} &= \frac{35}{64} e_3 e_1^3-\frac{75}{64} e_2 e_3 e_1+\frac{33 e_3^2}{64} ~,\nonumber \\
  \hat{f}_{3,2} &= -\frac{15015 e_3 e_1^6}{4096}+\frac{38115 e_2 e_3 e_1^4}{2048}-\frac{29925 e_3^2 e_1^3}{2048}-\frac{93555 e_2^2 e_3 e_1^2}{4096}
            +\frac{46095 e_2 e_3^2 e_1}{2048}+\frac{3955 e_2^3 e_3}{1024}-\frac{14595 e_3^3}{4096} ~,\nonumber \\
  \hat{f}_{3,3} &= \frac{8083075 e_3 e_1^9}{131072}-\frac{65090025 e_2 e_3 e_1^7}{131072}+\frac{56531475 e_3^2 e_1^6}{131072}+\frac{172297125 e_2^2 e_3 e_1^5}{131072}
            -\frac{122207085 e_2 e_3^2 e_1^4}{65536} \nonumber \\
          &\phantom{abc}+\frac{80627085 e_3^3 e_1^3}{131072}-\frac{165840675 e_2^3 e_3 e_1^3}{131072}+\frac{247328235 e_2^2 e_3^2 e_1^2}{131072}+\frac{10481625 e_2^4 e_3 e_1}{32768}\nonumber \\
          &\phantom{abc} -\frac{97970985 e_2 e_3^3 e_1}{131072}+\frac{7978285 e_3^4}{131072}-\frac{8327655 e_2^3 e_3^2}{32768}~,\nonumber \\
  \hat{f}_{3,4} &= -\frac{32534376875 e_3 e_1^{12}}{16777216}+\frac{89794880175 e_2 e_3 e_1^{10}}{4194304}-\frac{81307651425 e_3^2 e_1^9}{4194304}
            -\frac{737879667525 e_2^2 e_3 e_1^8}{8388608} \nonumber \\
          &\phantom{abc}+\frac{588019457025 e_2 e_3^2 e_1^7}{4194304}+\frac{692291124525 e_2^3 e_3 e_1^6}{4194304}-\frac{455097968325 e_3^3 e_1^6}{8388608}
            -\frac{1379583079875 e_2^2 e_3^2 e_1^5}{4194304}\nonumber \\
          &\phantom{abc}+\frac{858144412125 e_2 e_3^3 e_1^4}{4194304}-\frac{2322086641875 e_2^4 e_3 e_1^4}{16777216}
            +\frac{1157389608375 e_2^3 e_3^2 e_1^3}{4194304}-\frac{162384146925 e_3^4 e_1^3}{4194304}\nonumber \\
          &\phantom{abc}+\frac{87699236625 e_2^5 e_3 e_1^2}{2097152}-\frac{1481627572425 e_2^2 e_3^3 e_1^2}{8388608}+\frac{163545349275 e_2 e_3^4 e_1}{4194304}
            -\frac{62462324925 e_2^4 e_3^2 e_1}{1048576}\nonumber \\
          &\phantom{abc}+\frac{41443427025 e_2^3 e_3^3}{2097152}-\frac{2103075975 e_2^6 e_3}{1048576}-\frac{32314471875 e_3^5}{16777216} \nonumber ~, \\
  \hat{f}_{3,5} &= \frac{52647128659125 e_3 e_1^{15}}{536870912}-\frac{738758095700625 e_2 e_3 e_1^{13}}{536870912}+\frac{684295548811875 e_3^2 e_1^{12}}{536870912}\nonumber \\
          &\phantom{abc}-\frac{1741727866003125 e_2 e_3^2 e_1^{10}}{134217728}+\frac{1447231892030925 e_3^3 e_1^9}{268435456}
            -\frac{5837805914565625 e_2^3 e_3 e_1^9}{268435456}\nonumber \\
          &\phantom{abc}+\frac{13104215838338625 e_2^2 e_3^2 e_1^8}{268435456}+\frac{17576000871053625 e_2^4 e_3 e_1^7}{536870912}
            -\frac{9503237789320275 e_2 e_3^3 e_1^7}{268435456}\nonumber \\
          &\phantom{abc}+\frac{2209625052609975 e_3^4 e_1^6}{268435456}-\frac{11160180446790375 e_2^3 e_3^2 e_1^6}{134217728}
            +\frac{20033877534345675 e_2^2 e_3^3 e_1^5}{268435456}\nonumber \\
          &\phantom{abc}-\frac{13477372127344125 e_2^5 e_3 e_1^5}{536870912}+\frac{33627317776927875 e_2^4 e_3^2 e_1^4}{536870912}
            -\frac{3697210384993125 e_2 e_3^4 e_1^4}{134217728}\nonumber \\
          &\phantom{abc}+\frac{1840228057292625 e_3^5 e_1^3}{536870912}+\frac{564221631348375 e_2^6 e_3 e_1^3}{67108864}
            -\frac{14908243313588625 e_2^3 e_3^3 e_1^3}{268435456}\nonumber \\
          &\phantom{abc}+\frac{5572189269021375 e_2^2 e_3^4 e_1^2}{268435456}-\frac{1126515991726125 e_2^5 e_3^2 e_1^2}{67108864}
            +\frac{702424017169875 e_2^4 e_3^3 e_1}{67108864}\nonumber \\
          &\phantom{abc}-\frac{27007478623125 e_2^7 e_3 e_1}{33554432}-\frac{1584673523658225 e_2 e_3^5 e_1}{536870912}
            +\frac{23592916722375 e_2^6 e_3^2}{33554432} \nonumber \\
          &\phantom{abc}+\frac{2063427784543125 e_2^2 e_3 e_1^{11}}{268435456}+\frac{52435195988175 e_3^6}{536870912}
            -\frac{133329976054275 e_2^3 e_3^4}{67108864} ~.
\ea
\paragraph{$n=4$}
\ba
  \la{B.14}
  \hat{f}_{4,0} &= -\frac{3}{4}  e_1 e_4~,\nonumber \\
  \hat{f}_{4,1} &= \frac{315}{128} e_4 e_1^4-\frac{945}{128} e_2 e_4 e_1^2+\frac{615}{128} e_3 e_4 e_1+\frac{75}{32} e_2^2 e_4-\frac{159 e_4^2}{64}~,\nonumber \\
  \hat{f}_{4,2} &=-\frac{225225 e_4 e_1^7}{8192}+\frac{675675 e_2 e_4 e_1^5}{4096}-\frac{557865 e_3 e_4 e_1^4}{4096}
            +\frac{111825 e_4^2 e_1^3}{1024}-\frac{2248785 e_2^2 e_4 e_1^3}{8192}+\frac{1335285 e_2 e_3 e_4 e_1^2}{4096}\nonumber \\
          &\phantom{abc}-\frac{41895}{256} e_2 e_4^2 e_1+\frac{222705 e_2^3 e_4 e_1}{2048}-\frac{655305 e_3^2 e_4 e_1}{8192}
            +\frac{109095 e_3 e_4^2}{2048}-\frac{40845}{512} e_2^2 e_3 e_4~,\nonumber \\
  \hat{f}_{4,3} &=\frac{169744575 e_4 e_1^{10}}{262144}-\frac{1527701175 e_2 e_4 e_1^8}{262144}+\frac{1351575225 e_3 e_4 e_1^7}{262144}
            +\frac{4751571825 e_2^2 e_4 e_1^6}{262144}\nonumber \\
          &\phantom{abc}-\frac{3545266725 e_2 e_3 e_4 e_1^5}{131072}+\frac{1267160895 e_2 e_4^2 e_1^4}{65536}+\frac{2514637125 e_3^2 e_4 e_1^4}{262144}
            -\frac{5933552625 e_2^3 e_4 e_1^4}{262144}\nonumber \\
          &\phantom{abc}+\frac{10111025925 e_2^2 e_3 e_4 e_1^3}{262144}-\frac{104757345 e_3 e_4^2 e_1^3}{8192}+\frac{125697285 e_4^3 e_1^2}{32768}
            +\frac{316891575 e_2^4 e_4 e_1^2}{32768}\nonumber \\
          &\phantom{abc}-\frac{5013144675 e_2 e_3^2 e_4 e_1^2}{262144}+\frac{250441065 e_2 e_3 e_4^2 e_1}{16384}+\frac{659630475 e_3^3 e_4 e_1}{262144}
            -\frac{837819675 e_2^3 e_3 e_4 e_1}{65536}\nonumber \\
          &\phantom{abc}+\frac{247868775 e_2^2 e_3^2 e_4}{65536}-\frac{10481625 e_2^5 e_4}{16384}-\frac{62462295 e_2 e_4^3}{32768}-\frac{246043245 e_3^2 e_4^2}{131072} \nonumber \\
          &\phantom{abc}-\frac{591666075 e_4^2 e_1^6}{131072}-\frac{2529529695 e_2^2 e_4^2 e_1^2}{131072}+ \frac{83600685 e_2^3 e_4^2}{32768} ~.
\ea
\paragraph{$n=5,6,7,8,9$}
\la{sec:n=5-6-7}
\ba
  \la{B.15}
  \hat{f}_{5,0} &=\frac{3 e_2 e_5}{2}-\frac{15}{8} e_1^2 e_5~,\nonumber \\
  \hat{f}_{5,1} &=\frac{3465}{256} e_5 e_1^5-\frac{13545}{256} e_2 e_5 e_1^3+\frac{9975}{256} e_3 e_5 e_1^2+\frac{2415}{64} e_2^2 e_5 e_1
            -\frac{3255}{128} e_4 e_5 e_1+\frac{1347 e_5^2}{128}-\frac{1515}{64} e_2 e_3 e_5 ~,\nonumber \\
  \hat{f}_{6,0} &=-\frac{105}{16}  e_6 e_1^3+\frac{45}{4} e_2 e_6 e_1-\frac{9 e_3 e_6}{2}~,\nonumber \\
  \hat{f}_{6,1} &=\frac{45045}{512} e_6 e_1^6-\frac{218295}{512} e_2 e_6 e_1^4+\frac{171045}{512} e_3 e_6 e_1^3+\frac{31185}{64} e_2^2 e_6 e_1^2
            -\frac{62685}{256} e_4 e_6 e_1^2-\frac{59535}{128} e_2 e_3 e_6 e_1\nonumber \\
          &\phantom{abc}+\frac{39285}{256} e_5 e_6 e_1-\frac{2415}{32} e_2^3 e_6+\frac{4545}{64} e_3^2 e_6+\frac{9315}{64} e_2 e_4 e_6-\frac{2385 e_6^2}{32}~,\nonumber \\
  \hat{f}_{7,0} &=-\frac{945}{32} e_7 e_1^4+\frac{315}{4} e_2 e_7 e_1^2-45 e_3 e_7 e_1-\frac{45}{2} e_2^2 e_7+18 e_4 e_7\nonumber \\
  \hat{f}_{7,1} &=\frac{675675 e_7 e_1^7}{1024}-\frac{3918915 e_2 e_7 e_1^5}{1024}+\frac{3191265 e_3 e_7 e_1^4}{1024}
            +\frac{779625}{128} e_2^2 e_7 e_1^3-\frac{1248345}{512} e_4 e_7 e_1^3-\frac{898695}{128} e_2 e_3 e_7 e_1^2\nonumber \\
          &\phantom{abc}+\frac{901845}{512} e_5 e_7 e_1^2-\frac{146475}{64} e_2^3 e_7 e_1+\frac{52605}{32} e_3^2 e_7 e_1+\frac{428715}{128} e_2 e_4 e_7 e_1-\frac{141705}{128} e_6 e_7 e_1\nonumber \\
          &\phantom{abc}+\frac{55845 e_7^2}{128}+\frac{103005}{64} e_2^2 e_3 e_7-\frac{64305}{64} e_3 e_4 e_7-\frac{132435}{128} e_2 e_5 e_7~,\nonumber \\
  \hat{f}_{8,0} &=-\frac{10395}{64}  e_8 e_1^5+\frac{4725}{8} e_2 e_8 e_1^3-\frac{1575}{4} e_3 e_8 e_1^2-\frac{1575}{4} e_2^2 e_8 e_1
            +225 e_4 e_8 e_1+225 e_2 e_3 e_8-90 e_5 e_8 ~,\nonumber \\
  \hat{f}_{8,1} &=\frac{11486475 e_8 e_1^8}{2048}-\frac{77702625 e_2 e_8 e_1^6}{2048}+\frac{64999935 e_3 e_8 e_1^5}{2048}+\frac{39864825}{512} e_2^2 e_8 e_1^4-\frac{26496855 e_4 e_8 e_1^4}{1024}\nonumber \\
          &\phantom{abc}-\frac{25623675}{256} e_2 e_3 e_8 e_1^3+\frac{20600055 e_5 e_8 e_1^3}{1024}-\frac{6288975}{128} e_2^3 e_8 e_1^2+\frac{3642975}{128} e_3^2 e_8 e_1^2+\frac{7396515}{128} e_2 e_4 e_8 e_1^2\nonumber \\
          &\phantom{abc}-\frac{1848735}{128} e_6 e_8 e_1^2+\frac{7158375}{128} e_2^2 e_3 e_8 e_1-\frac{427455}{16} e_3 e_4 e_8 e_1-\frac{7017885}{256} e_2 e_5 e_8 e_1+\frac{2263095}{256} e_7 e_8 e_1\nonumber \\
          &\phantom{abc}+\frac{146475}{32} e_2^4 e_8-\frac{414225}{32} e_2 e_3^2 e_8+\frac{64305}{16} e_4^2 e_8-\frac{840735}{64} e_2^2 e_4 e_8+\frac{1040355}{128} e_3 e_5 e_8\nonumber \\
          &\phantom{abc}+\frac{269505}{32} e_2 e_6 e_8-\frac{545175 e_8^2}{128}~,\nonumber \\
  \hat{f}_{9,0} &= -\frac{135135}{128}  e_9 e_1^6+\frac{155925}{32} e_2 e_9 e_1^4-\frac{14175}{4} e_3 e_9 e_1^3-\frac{42525}{8} e_2^2 e_9 e_1^2
            +\frac{4725}{2} e_4 e_9 e_1^2\nonumber \\
          & \phantom{abc}+4725 e_2 e_3 e_9 e_1-1350 e_5 e_9 e_1+\frac{1575}{2} e_2^3 e_9-675 e_3^2 e_9-1350 e_2 e_4 e_9+540 e_6 e_9 ~.
\ea

\paragraph{\underline{First subleading order $\delta{\omega}_{n,g}$}}\la{app:sec:first-subl-order}
\paragraph{$n=1$}
\ba
  \la{B.16}
  \delta f_{1,1} &= \frac{5 e_1}{128} ~,\nonumber \\
  \delta f_{1,2} &=-\frac{483 e_1^4}{8192} ~,\nonumber \\
  \delta f_{1,3} &=\frac{137137 e_1^7}{262144} ~,\nonumber \\
  \delta f_{1,4} &=-\frac{370204835 e_1^{10}}{33554432} ~,\nonumber \\
  \delta f_{1,5} &=\frac{448974400875 e_1^{13}}{1073741824} ~.
\ea
\paragraph{$n=2$}
\ba
  \la{B.17}
  \delta f_{2,1} &=\frac{5 e_1^3}{256}+\frac{3 e_2 e_1}{256} ~,\nonumber \\
  \delta f_{2,2} &=-\frac{1155 e_1^6}{16384}+\frac{819 e_2 e_1^4}{8192}+\frac{5817 e_2^2 e_1^2}{16384}-\frac{833 e_2^3}{4096} ~,\nonumber \\
  \delta f_{2,3} &=\frac{425425 e_1^9}{524288}-\frac{1396395 e_2 e_1^7}{524288}-\frac{3252249 e_2^2 e_1^5}{524288}+\frac{12156375 e_2^3 e_1^3}{524288}
                 -\frac{1591065 e_2^4 e_1}{131072} ~,\nonumber \\
  \delta f_{2,4} &=-\frac{1301375075 e_1^{12}}{67108864}+\frac{1663496835 e_2 e_1^{10}}{16777216}+\frac{5426976555 e_2^2 e_1^8}{33554432}
                 -\frac{27126714615 e_2^3 e_1^6}{16777216}\nonumber \\
               &\phantom{abc}+\frac{184265806725 e_2^4 e_1^4}{67108864}-\frac{11560844295 e_2^5 e_1^2}{8388608}+\frac{427732305 e_2^6}{4194304} ~.
\ea
\paragraph{$n=3$}
\ba
  \la{B.18}
  \delta f_{3,0} &= -\frac{e_2}{16} ~,\nonumber \\
  \delta f_{3,1} &=\frac{35}{512} e_2 e_1^3+\frac{15}{256} e_3 e_1^2-\frac{75}{512} e_2^2 e_1+\frac{21 e_2 e_3}{512} ~,\nonumber \\
  \delta f_{3,2} &=-\frac{15015 e_2 e_1^6}{32768}-\frac{7623 e_3 e_1^5}{8192}+\frac{38115 e_2^2 e_1^4}{16384}+\frac{29925 e_2 e_3 e_1^3}{16384}-\frac{21945 e_3^2 e_1^2}{8192}-\frac{93555 e_2^3 e_1^2}{32768}\nonumber \\
               &\phantom{abc}+\frac{3255 e_2^2 e_3 e_1}{16384}+\frac{3955 e_2^4}{8192}+\frac{37485 e_2 e_3^2}{32768} ~,\nonumber \\
  \delta f_{3,3} &=\frac{8083075 e_2 e_1^9}{1048576}+\frac{13018005 e_3 e_1^8}{524288}-\frac{65090025 e_2^2 e_1^7}{1048576}-\frac{125450325 e_2 e_3 e_1^6}{1048576}+\frac{172297125 e_2^3 e_1^5}{1048576}\nonumber \\
               &\phantom{abc}+\frac{38657619 e_3^2 e_1^5}{262144}+\frac{18063045 e_2^2 e_3 e_1^4}{131072}-\frac{165840675 e_2^4 e_1^3}{1048576}-\frac{432151335 e_2 e_3^2 e_1^3}{1048576}+\frac{75661425 e_3^3 e_1^2}{524288}\nonumber \\
               &\phantom{abc}-\frac{10481625 e_2^3 e_3 e_1^2}{1048576}+\frac{10481625 e_2^5 e_1}{262144}+\frac{205872975 e_2^2 e_3^2 e_1}{1048576}-\frac{1963395 e_2^4 e_3}{262144}-\frac{65346575 e_2 e_3^3}{1048576} ~.
\ea
\paragraph{$n=4,5$}
\ba
  \la{B.19}
  \delta f_{4,0} &=\frac{3 e_4}{16}-\frac{3 e_1 e_3}{32} ~,\nonumber \\
  \delta f_{4,1} &=\frac{315 e_3 e_1^4}{1024}-\frac{945 e_2 e_3 e_1^2}{1024}+\frac{615 e_3^2 e_1}{1024}+\frac{75}{512} e_2 e_4 e_1
                 +\frac{75}{256} e_2^2 e_3-\frac{111 e_3 e_4}{256} ~,\nonumber \\
  \delta f_{4,2} &=-\frac{225225 e_3 e_1^7}{65536}-\frac{153153 e_4 e_1^6}{32768}+\frac{675675 e_2 e_3 e_1^5}{32768}+\frac{183645 e_2 e_4 e_1^4}{8192}
                 -\frac{557865 e_3^2 e_1^4}{32768}-\frac{31815 e_3 e_4 e_1^3}{8192}~,\nonumber \\
               &\phantom{abc}-\frac{2248785 e_2^2 e_3 e_1^3}{65536}+\frac{1335285 e_2 e_3^2 e_1^2}{32768}+\frac{106785 e_4^2 e_1^2}{8192}
                 -\frac{823095 e_2^2 e_4 e_1^2}{32768}+\frac{222705 e_2^3 e_3 e_1}{16384}~,\nonumber \\
               &\phantom{abc}+\frac{50295 e_2 e_3 e_4 e_1}{16384}-\frac{655305 e_3^3 e_1}{65536}+\frac{30975 e_2^3 e_4}{8192}+\frac{105735 e_3^2 e_4}{32768}
                 -\frac{40845 e_2^2 e_3^2}{4096}-\frac{61635 e_2 e_4^2}{8192} ~,\nonumber \\
  \delta f_{5,0} &=-\frac{15}{64}  e_4 e_1^2+\frac{21 e_5 e_1}{32}+\frac{3 e_2 e_4}{16} ~,\nonumber \\
  \delta f_{5,1} &=\frac{3465 e_4 e_1^5}{2048}-\frac{315}{256} e_5 e_1^4-\frac{13545 e_2 e_4 e_1^3}{2048}+\frac{9975 e_3 e_4 e_1^2}{2048}
                 +\frac{4305 e_2 e_5 e_1^2}{1024}+\frac{2415}{512} e_2^2 e_4 e_1\nonumber \\
               &\phantom{abc}-\frac{3255 e_4^2 e_1}{1024}-\frac{3225 e_3 e_5 e_1}{1024}-\frac{1515}{512} e_2 e_3 e_4-\frac{375}{256} e_2^2 e_5+\frac{3123 e_4 e_5}{1024} ~.
\ea

\section{The correlation function \texorpdfstring{$\vev{:{\rm tr} M:\mathcal{W}^{n}}$}{MWn}}
\la{app:sec:corr-funct-vev}

In the main text, to prove (\ref{7.21}), we exploited the fact that $\vev{:\tr M: \mc W^{n}}$ has no higher genus corrections beyond the leading order.
This can be easily proved by starting from the following splitting of $M$ in the $U(N)$ theory
\ba
  \la{C.1}
  M &= \tilde{M} + \frac{m}{N} ~,&& \tilde{M} = M - \frac{1}{N}\tr M ~,&& m = \tr M ~,
\ea
where $\tilde{M}$ is the traceless part. The matrix model partition function becomes
\ba
  \la{C.2}
  Z &= \int_{-\infty}^{\infty}dm \int d \tilde{M} \, \delta(\tr \tilde{M}) \exp\left(-\frac{N}{2} \tr \tilde{M}^{2} - \frac{m^{2}}{2}\right) ~.
\ea
For the Wilson loop operator, the splitting (\ref{C.1}) implies
\ba
  \tr \exp\left(\frac{\sqrt\lambda}{2} M\right) &= \exp\left(\frac{\sqrt{\lambda}}{2N} m\right)\tr \exp\left(\frac{\sqrt{\lambda}}{2} \tilde{M}\right) ~.
\ea
As a result the expectation value of $n$ coincident Wilson loops takes the form 
\ba
  \la{C.4}
  \vev{\mathcal{W}^{n}} &=  \vev{\mathcal{W}^{n}}_{\rm traceless}\, \int_{-\infty}^{\infty} d m \,  \exp\left(-\frac{m^{2}}{2} + \frac{n\lambda}{2N} m\right), \notag \\
\vev{\mc W^{n}}_{\rm traceless} &=   \int d\tilde{M}\,\delta(\tr \tilde{M})\left[\tr \exp\left(\frac{\lambda}{2} \tilde{M}\right)\right]^{n}\exp\left(-\frac{N}{2}\tr \tilde{M}^{2}\right) ~.
\ea
In the case of  $\vev{m \mathcal{W}^{n}}$, we obtain the same integral for $\tilde{M}$ with an extra insertion of $m$ in the $m$-integral. As a result, the ``traceless'' part $\vev{\mc W^{n}}_{\rm traceless}$
cancels and we obtain 
\ba
  \la{C.5}
  \frac{\vev{m \mathcal{W}^{n}}}{\vev{\mathcal{W}^{n}}} &= \frac{\int_{-\infty}^{\infty} d m \, m  \exp\left(-\frac{m^{2}}{2} + \frac{n\sqrt{\lambda}}{2N} m\right)}
  {\int_{-\infty}^{\infty} d m \,  \exp\left(-\frac{m^{2}}{2} + \frac{n\sqrt{\lambda}}{2N} m\right)} = \frac{n\sqrt{\lambda}}{2N}~.
\ea
This is just the leading order result obtained in \eqref{7.13} and specialized to $J=1$. The above discussion shows that it is in fact exact.

\bibliography{BT-Biblio}
\bibliographystyle{JHEP}

\end{document}